\documentclass[aps,prd,twocolumn,nofootinbib]{revtex4-2}

\usepackage{amsmath,amssymb,mathtools}
\usepackage{bm}
\usepackage{booktabs}
\usepackage{graphicx}
\usepackage{tikz}
\usepackage{placeins}

\usetikzlibrary{arrows.meta,positioning,fit,backgrounds,%
decorations.pathmorphing,decorations.markings}
\tikzset{fermion/.style={postaction={decorate},
  decoration={markings,mark=at position 0.58 with {\arrow{Latex[length=2mm]}}}}}
\usepackage{hyperref}
\hypersetup{colorlinks=true,citecolor=blue,linkcolor=blue,urlcolor=blue}
\newcommand{\bsig}{\boldsymbol{\sigma}}
\newcommand{\bq}{\mathbf q}
\newcommand{\bP}{\mathbf P}
\newcommand{\bp}{\mathbf p}
\newcommand{\br}{\mathbf r}
\newcommand{\bv}{\mathbf v}
\newcommand{\rhat}{\hat{\mathbf r}}
\newcommand{\dd}{\mathrm d}
\newcommand{\Lag}{\mathcal L}
\newcommand{\Mcal}{\mathcal M}
\newcommand{\Vt}{\widetilde V}

\newcommand{\bL}{\boldsymbol{L}}

\begin{document}

\title{Spin-dependent fermion potentials from mixed tensor couplings of massive spin-1 and spin-2 bosons}

\author{D. Khadka}
\author{V. V. Flambaum}\email{v.flambaum@unsw.edu.au}
\affiliation{School of Physics, University of New South Wales, Sydney 2052, Australia}

\begin{abstract}
Searches for weak spin-dependent forces are commonly interpreted in terms of a basis of sixteen rotationally invariant two-fermion potentials $V_{1-16}$, including potentials which violate parity (P) and time-reversal (T) invariance.  We derive finite-range coordinate-space potentials generated by a massive spin-1 boson with vector, axial-vector, tensor, and pseudotensor couplings to Dirac fermions.  In addition to the familiar vector--vector, vector--axial-vector, and axial-vector--axial-vector interactions, we obtain the mixed vector--tensor, vector--pseudotensor, axial-vector--tensor, and axial-vector--pseudotensor potentials, together with tensor--tensor, tensor--pseudotensor, and pseudotensor--pseudotensor terms. 
We retain the contact terms needed in atomic-scale applications and give compact expressions that include both orderings of inequivalent fermion vertices.  We then extend the analysis to a massive spin-2 mediator described by a symmetric Fierz--Pauli field.  Besides the conventional coupling to the conserved fermion energy--momentum tensor, we consider a symmetric axial-tensor current.  Its nonconservation for massive fermions makes the longitudinal spin-2 helicities contribute to axial-tensor--axial-tensor exchange.  The leading minimal spin-2 interactions are either $P,T$ even or $P$ odd and $T$ even; in particular, they do not generate a $T$-odd potential.  Heavy-mediator contact limits and the distinction between the $M\to0$ massive-spin-2 limit and a genuine massless graviton are also discussed.
We map the resulting interactions onto the sixteen-potential Dobrescu--Mocioiu basis and use this mapping to translate existing experimental constraints into numerical limits on spin-1 and spin-2 coupling products in the small- and large-mediator-mass regimes, together with representative limits at finite mediator masses. The derived relations between the boson-exchange coupling constants and the coefficients of the basis potentials also allow published exclusion curves to be rescaled to arbitrary mediator masses.

\end{abstract}

\maketitle

\section{Introduction}
\label{sec:intro}

Weak, long-range interactions between fermions provide a broad class of probes of physics beyond the Standard Model.  The classic scalar and pseudoscalar exchange potentials were developed in Ref.~\cite{MoodyWilczek}, while a systematic symmetry classification of spin-dependent interactions was given by Dobrescu and Mocioiu~\cite{DobrescuMocioiu}.  Coordinate-space forms appropriate both to macroscopic and atomic-scale experiments, including contact terms, were subsequently developed in Refs.~\cite{Fadeev2019,Cong2025}.  The latter review provides a comprehensive account of present experimental searches.

\begin{figure}[h]
\centering
\includegraphics[width=0.25\textwidth]{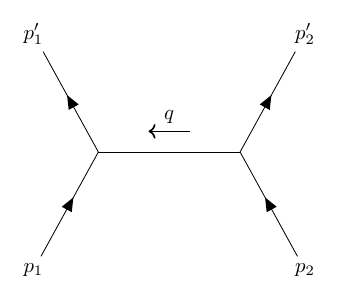} 
\caption{Elastic scattering of two fermions via the exchange of a boson}
\label{fig:4}
\end{figure}

For two spin-$1/2$ particles, rotational invariance permits sixteen independent nonrelativistic operator structures when scalar coefficient functions of the two independent momenta are factored out~\cite{DobrescuMocioiu}. 

The first purpose of this paper is to consider a massive neutral spin-1 field with the combined interaction with a fermion
\begin{equation}
\begin{split}
\Lag_{1}={}& Z'_\mu\sum_f \bar f\gamma^\mu
 \left(g_V^f+g_A^f\gamma_5\right)f \\
&+ Z'_{\mu\nu}\sum_f \bar f\sigma^{\mu\nu}
 \left(g_T^f+i g_{\widetilde T}^f\gamma_5\right)f,
\end{split}
\label{eq:Lspin1intro}
\end{equation}
where $Z'_{\mu\nu}=\partial_\mu Z'_\nu-\partial_\nu Z'_\mu$.
The second line of Eq. (\ref{eq:Lspin1intro})  is analogous to the electromagnetic dipole coupling of a fermion's anomalous magnetic and electric dipole moments to the field-strength tensor;
$g_T$ and $g_{\widetilde T}$ therefore have mass dimension $-1$. 
The explicit factor $i$ in the pseudotensor term makes the operator Hermitian for real diagonal couplings. This interaction produces interference terms between the vector/axial and tensor/pseudotensor vertices.
The vector-tensor potential
have been considered for the meson-exchange models of the nucleon-nucleon interaction  \cite{Machleidt}. The vector-tensor potential derived in our work is  generalized to arbitrary fermion masses and independent couplings at the two vertices. The contact terms are also explicitly retained in our work.

\begin{figure}[t]
\centering
\includegraphics[width=\columnwidth]{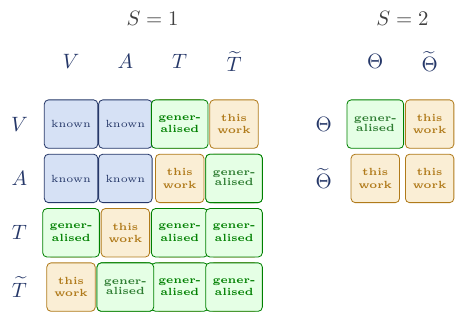} 
\caption{Interaction between two fermions mediated by spin-1 and spin-2 bosons.
Rows and columns label the vertex type on each fermion line. For the massive
spin-1 mediator of Eq.~\eqref{eq:Lspin1intro} these are the vector ($V$), axial
vector ($A$), tensor ($T$) and pseudotensor ($\widetilde T$) vertices; for the
massive spin-2 mediator of Eq.~\eqref{eq:Lspin2} they are the
energy--momentum ($\Theta$) and axial-tensor ($\widetilde\Theta$) currents.
Only $g_Vg_V$, $g_Ag_V$ and $g_Ag_A$ are standard Proca-exchange
potentials. The spin-1 channels involving at least one field-strength
(dipole) vertex are considered in Sec.~\ref{sec:spin1}.  The vector-tensor $(VT)$ interaction derived by Machleidt \textit{et al.}~\cite{Machleidt} for nucleon--nucleon interactions mediated by a meson exchange is generalized here to arbitrary fermion masses and independent couplings at the two vertices. The contact terms are also explicitly retained in our work. The time-reversal-odd, parity-even (TOPE) axial--pseudotensor interaction considered previously
\cite{Khriplovich1991,Conti1992,Porsev1994}, is generalized here to finite mediator mass. The tensor--tensor $g_Tg_T$, pseudotensor--tensor $g_{\widetilde T}g_T$, and pseudotensor--pseudotensor $g_{\widetilde T}g_{\widetilde T}$  potentials, derived previously for the massless case \cite{Fadeev2019}, are also generalized here to finite mediator mass. In the spin-2
sector, the conventional $\Theta\Theta$ interaction has been studied
previously \cite{Kang2021} and is re-derived and generalized here, while the mixed $\Theta\widetilde\Theta$ and $\widetilde\Theta\widetilde\Theta$ channels are derived in
Sec.~\ref{sec:spin2}.}
\label{f:matrix}
\end{figure}

The second purpose is to extend the same potential framework to a massive spin-2 mediator.  Massive spin-2 fields and their couplings to matter have been studied extensively in the Fierz--Pauli and effective-field-theory descriptions~\cite{HanLykkenZhang,Koenigstein,Hinterbichler,Kang2021,Voronchikhin1,Voronchikhin2}. 
The standard fermion coupling is to a symmetric conserved energy--momentum tensor. 
Here we also consider its symmetric axial counterpart, a Lorentz structure previously used as an interpolating current for negative-parity $J^{PC}=2^{--}$ tensor mesons in QCD \cite{AlievBilmisYang2018,AlievBarakatSavci2018}.
This operator is distinct from the antisymmetric Dirac tensor $\sigma^{\mu\nu}$ in Eq.~\eqref{eq:Lspin1intro}. 
We derive the leading nonrelativistic potentials and determine their $P$ and $T$ properties. Figure~\ref{f:matrix} summarizes which of the resulting boson exchange potentials are already available in the literature and which are considered here.
 We also  map the boson exchange interactions onto the sixteen-potential Dobrescu--Mocioiu basis presented in Ref.~\cite{DobrescuMocioiu}. We use this mapping to obtain numerical translations of existing limits for both spin-1 and spin-2 coupling products in the small- and large-mediator-mass regimes, together with representative limits at finite mediator masses.The relations derived here between the interaction constants of the boson-exchange potentials $g_i$ and the coefficients of the basis potentials $f_i$ allow published exclusion plots for $f_i$ to be rescaled to $g_i g_k$ for arbitrary mediator masses. We therefore do not reproduce these exclusion plots in the present paper.

We use natural units $\hbar=c=1$ and metric $\eta_{\mu\nu}=\mathrm{diag}(1,-1,-1,-1)$ throughout.

\section{Scattering formalism and useful Fourier kernels}
\label{sec:formalism}

Consider elastic scattering
\begin{equation}
 f_1(p_1)f_2(p_2)\to f_1(p'_1)f_2(p'_2),
\label{eq:process}
\end{equation}
with momentum transfer
\begin{equation}
 q=p'_1-p_1=p_2-p'_2,\qquad q^0\simeq0.
\label{eq:qdef}
\end{equation}
Throughout, $\bp_a$ denotes the \emph{average} three-momentum of
fermion $a$, and $\bv$ the relative velocity,
\begin{align}
 \bp_a&=\tfrac12\left(\bp_a^{\rm in}+\bp_a^{\rm out}\right),
\label{eq:pavdef}\\
 \bv&=\frac{\bp_1}{m_1}-\frac{\bp_2}{m_2}.
\label{eq:vdef}
\end{align}
In the center-of-mass frame it is also useful to write
\begin{equation}
 \bP\equiv\bp_1=-\bp_2=m_r\bv,
\label{eq:Preduced}
\end{equation}
where $m_r=m_1m_2/(m_1+m_2)$ is the reduced mass.
With relativistically normalized external spinors, $\bar uu\simeq2m$, the Born matching convention used below is
\begin{align}
 \Vt(\bq)&=-\frac{\Mcal_{\rm NR}(\bq)}{4m_1m_2},
\label{eq:Born}\\
 V(\br)&=\int\frac{\dd^3q}{(2\pi)^3}e^{i\bq\cdot\br}\,\Vt(\bq).
\label{eq:BornFT}
\end{align}
Changing the overall sign convention of the interaction Lagrangian changes the overall potential sign but not the relative operator structures or symmetry classifications.

It is useful to collect the radial distributions that occur repeatedly. We define
\begin{align}
Y_M(r)&=\frac{e^{-Mr}}{4\pi r},
\label{eq:Y}\\
F_M(r)&=-\frac{\dd Y_M}{\dd r}
=\frac{e^{-Mr}}{4\pi}\left(\frac1{r^2}+\frac Mr\right),
\label{eq:F}\\
D_M(\br)&=\delta^3(\br)-M^2Y_M(r),
\label{eq:D}
\end{align}
so that $D_M$ is the transform of $\bq^2/(\bq^2+M^2)$.  We also define
\begin{align}
H_{ij}(\br)&=\int\frac{\dd^3q}{(2\pi)^3}e^{i\bq\cdot\br}
\frac{q_iq_j}{\bq^2+M^2}
\nonumber\\
&=\frac{e^{-Mr}}{4\pi}\biggl[
\delta_{ij}\left(\frac1{r^3}+\frac M{r^2}\right)
\nonumber\\
&\qquad-\hat r_i\hat r_j
\left(\frac3{r^3}+\frac{3M}{r^2}+\frac{M^2}{r}\right)\biggr]
\nonumber\\
&\quad+\frac13\delta_{ij}\delta^3(\br),
\label{eq:H}
\end{align}
and
\begin{equation}
Q_{12}(\br)=(\bsig_1\cdot\bsig_2)D_M(\br)
-\sigma_{1i}\sigma_{2j}H_{ij}(\br).
\label{eq:Q}
\end{equation}
Thus $Q_{12}$ is the transform of
$[(\bq\times\bsig_1)\cdot(\bq\times\bsig_2)]/(\bq^2+M^2)$.
For products containing momentum operators we use the Hermitian prescription
\begin{equation}
 p_i f(\br)\ \longrightarrow\ \frac12\{p_i,f(\br)\},
\label{eq:symm}
\end{equation}
where $\{A,B\}=AB+BA$.

\section{Massive spin-1 mediator with vector and dipole couplings}
\label{sec:spin1}

\subsection{Relativistic vertices and transversality}

For fermion species $a$, the two types of spin-1 vertices in Eq.~\eqref{eq:Lspin1intro} are
\begin{align}
\Gamma_{V/A,a}^{\mu}&=\gamma^\mu(g_V^a+g_A^a\gamma_5),
\label{eq:VAvertex}\\
\Gamma_{T/\widetilde T,a}^{\nu}&=-2iq_\mu\sigma^{\mu\nu}
(g_T^a+i g_{\widetilde T}^a\gamma_5),
\label{eq:Tvertex}
\end{align}
where the sign of $q$ is reversed at the second fermion line.  The Proca propagator is
\begin{equation}
D_{\mu\nu}(q)=\frac{-i}{q^2-M^2+i0}
\left(\eta_{\mu\nu}-\frac{q_\mu q_\nu}{M^2}\right).
\label{eq:Proca}
\end{equation}
The vector current is conserved on shell.  In addition, each field-strength tensor vertex is exactly transverse in its remaining vector index,
\begin{equation}
q_\nu q_\mu\bar u'\sigma^{\mu\nu}(1,\gamma_5)u=0,
\label{eq:dipoletransverse}
\end{equation}
because $q_\mu q_\nu$ is symmetric and $\sigma^{\mu\nu}$ is antisymmetric.  Thus the longitudinal Proca term drops out from every amplitude containing at least one tensor or pseudotensor field-strength vertex.  Notice that Eq.~\eqref{eq:dipoletransverse} does \emph{not} mean that the rank-two Dirac tensor current itself is conserved; in fact
\begin{equation}
q_\mu\bar u'\sigma^{\mu\nu}u
=i\bar u'\left(2m\gamma^\nu-K^\nu\right)u,
\label{eq:sigmadiv}
\end{equation}
with $K=p'+p$.

The leading nonrelativistic bilinears required below are listed in Appendix~\ref{app:NR}.  We now give compact coordinate-space results.  They include both possible orderings of inequivalent vertices, $1\leftrightarrow2$.

\subsection{Vector--tensor interaction}

The complete vector--tensor result, through the order retained here, is
\begin{align}
V_{VT}(\br)={}&-
\left(\frac{g_V^1g_T^2}{m_2}+\frac{g_T^1g_V^2}{m_1}\right)D_M
\nonumber\\
&-\left(\frac{g_V^1g_T^2}{m_1}+\frac{g_T^1g_V^2}{m_2}\right)Q_{12}
\nonumber\\
&+g_V^1g_T^2\left\{\bsig_2\cdot(\bv\times\rhat),F_M\right\}
\nonumber\\
&+g_T^1g_V^2\left\{\bsig_1\cdot(\bv\times\rhat),F_M\right\}.
\label{eq:VVTcompact}
\end{align}
The four terms contain a momentum-suppressed central interaction $V_1$, spin--spin/tensor terms $V_2+V_3$, and the $V_{4,5}$ spin--orbit structures.  (The potentials $V_i$ are
defined in Ref.~\cite{Cong2025}; see also Tables~\ref{tab:mapping} and
\ref{tab:RMPpotentials}.)  All are $P$ and $T$ even.

\subsection{Vector--pseudotensor interaction}

For a vector vertex and a pseudotensor dipole vertex we obtain
\begin{align}
V_{V\widetilde T}(\br)={}&
2g_V^1g_{\widetilde T}^2(\bsig_2\cdot\rhat)F_M
\nonumber\\
&-2g_{\widetilde T}^1g_V^2(\bsig_1\cdot\rhat)F_M
\nonumber\\
&-\frac{g_V^1g_{\widetilde T}^2}{2m_1m_2}
\sigma_{2j}\left\{(\bp_2\times\bsig_1)_i,H_{ij}\right\}
\nonumber\\
&-\frac{g_{\widetilde T}^1g_V^2}{2m_1m_2}
\sigma_{1j}\left\{(\bp_1\times\bsig_2)_i,H_{ij}\right\}.
\label{eq:VVPTcompact}
\end{align}
The first line is the familiar monopole--dipole $V_{9,10}$ structure, while the second contains the $q^2V_{14}\pm V_{15}$ combination.  The interaction is odd under both $P$ and $T$.

\subsection{Axial-vector--tensor interaction}

The axial-vector--tensor potential is
\begin{align}
V_{AT}(\br)={}&
-\frac{g_A^1g_T^2}{2m_1m_2}\left\{\bsig_1\cdot\bp_1,D_M\right\}
\nonumber\\
&-\frac{g_T^1g_A^2}{2m_1m_2}\left\{\bsig_2\cdot\bp_2,D_M\right\}
\nonumber\\
&-2\left(g_A^1g_T^2+g_T^1g_A^2\right)
(\bsig_1\times\bsig_2)\cdot\rhat\,F_M.
\label{eq:VATcompact}
\end{align}
It contains momentum-suppressed $V_{12,13}$ terms and the $V_{11}$ spin--spin structure.  Both are $P$ odd and $T$ even.

\subsection{Axial-vector--pseudotensor interaction and the TOPE potential}
\label{subsec:TOPE}

The axial-vector--pseudotensor cross term has an especially simple form,
\begin{align}
V_{A\widetilde T}(\br)={}&
 g_A^1g_{\widetilde T}^2
 \left\{\bsig_1\cdot\bv,(\bsig_2\cdot\rhat)F_M\right\}
\nonumber\\
&+g_{\widetilde T}^1g_A^2
 \left\{\bsig_2\cdot\bv,(\bsig_1\cdot\rhat)F_M\right\}.
\label{eq:VATpt}
\end{align}
It is a linear combination of the two $V_{6,7}$ operators,
\begin{equation}
(\bsig_1\cdot\bv)(\bsig_2\cdot\rhat)
\ \pm\ 
(\bsig_1\cdot\rhat)(\bsig_2\cdot\bv).
\label{eq:V67structure}
\end{equation}
Under parity, $\bv\to-\bv$ and $\rhat\to-\rhat$, whereas spin is an axial vector, so Eq.~\eqref{eq:VATpt} is $P$ even.  Under time reversal, $\bv\to-\bv$, $\bsig_a\to-\bsig_a$, and $\rhat\to\rhat$; consequently Eq.~\eqref{eq:VATpt} is $T$ odd.  Thus a single massive spin-1 boson with simultaneous axial-vector and pseudotensor couplings produces the TOPE interaction at tree level.

The $g_Ag_{\widetilde T}$ mechanism for a $P$-even, $T$-odd
(TOPE) interaction has been considered previously
\cite{Khriplovich1991,Conti1992,Porsev1994}. In particular,
Conti and Khriplovich derived the long-range electron--nucleon
interaction containing both independent TOPE structures. We show
below that their result is recovered from Eq.~\eqref{eq:VATpt} in the
long-range limit, while Eq.~\eqref{eq:VATpt} provides its finite-range
generalization.

For an electron--nucleon system, take particle $1=e$ and particle
$2=N$, and define
\begin{equation}
A=g_A^e g_{\widetilde T}^{N},
\qquad
B=g_{\widetilde T}^{e}g_A^N ,
\label{eq:ABTOPE}
\end{equation}
together with
\begin{align}
X&=
\left\{\bsig_e\cdot\bv,
(\bsig_N\cdot\rhat)F_M\right\},
\nonumber\\
Y&=
\left\{\bsig_N\cdot\bv,
(\bsig_e\cdot\rhat)F_M\right\}.
\label{eq:XYTOPE}
\end{align}
Equation~\eqref{eq:VATpt} can then be written as
\begin{equation}
V_{A\widetilde T}
=
AX+BY
=
\frac{A+B}{2}(X+Y)
+
\frac{B-A}{2}(Y-X).
\label{eq:TOPEdecomp}
\end{equation}

We compare this with the long-range TOPE interaction of Conti and Khriplovich given by 
\begin{equation}
\begin{aligned}
V=\frac{1}{4m_pm_r}
\Biggl\{
(\beta_{eN}+\beta_{Ne})\,i\Bigl[(\bsig_1\cdot\bp)(\bsig_2\cdot\bp)\frac1r
\\-\frac1r(\bsig_2\cdot\bp)(\bsig_1\cdot\bp)\Bigr]\\
-(\beta_{eN}-\beta_{Ne})\,\frac{(\bsig_1\times\bsig_2)\cdot\bL}{r^{3}}
\Biggr\}.
\label{eq:CK21}
\end{aligned}
\end{equation}
where $\beta_{eN}$ and $\beta_{Ne}$ are the TOPE electron-nucleon dimensionless constants. $m_p$ is not a dynamical fermion mass but the reference mass
introduced in Ref.~\cite{Conti1992} through the factor $1/(2m)$,
with $m$ conventionally fixed to the proton mass so that $\beta$
is dimensionless.

For $M\to0$, 
\begin{equation}
F_0(r)=\frac{1}{4\pi r^2}.
\end{equation}
Defining
\begin{equation}
D(r)=\frac{1}{4\pi r},
\end{equation}
we have
\begin{equation}
-\boldsymbol{\nabla}D=\rhat F_0,
\qquad
\hat r_iF_0=-i[p_i,D].
\label{eq:TOPEcommidentity}
\end{equation}
In the center-of-mass frame,
\begin{equation}
\bv=\frac{\bp}{m_r},
\qquad
m_r=\frac{m_em_N}{m_e+m_N},
\end{equation}
and, defining
\begin{equation}
P_e=\bsig_e\cdot\bp,
\qquad
P_N=\bsig_N\cdot\bp,
\end{equation}
Eq.~\eqref{eq:XYTOPE} gives
\begin{equation}
X=-\frac{i}{m_r}\{P_e,[P_N,D]\},
\qquad
Y=-\frac{i}{m_r}\{P_N,[P_e,D]\}.
\end{equation}
Since $[P_e,P_N]=0$, the symmetric combination becomes
\begin{align}
X+Y
&=
-\frac{2i}{m_r}[P_eP_N,D]
\nonumber\\
&=
-\frac{i}{2\pi m_r}
\left[
(\bsig_e\cdot\bp)(\bsig_N\cdot\bp),\frac1r
\right]
\label{eq:TOPEsym}
\end{align}
\begin{equation}
     \Rightarrow i\left[
(\bsig_e\cdot\bp)(\bsig_N\cdot\bp),\frac1r
\right]=-2\pi m_r(X+Y)
\label{eq:TOPEsym2}
\end{equation}

For the antisymmetric combination, define
\begin{equation}
A_j(\br)=\hat r_j F_0(r).
\label{eq:AjCK}
\end{equation}
Then we can write
\begin{equation}
Y-X=
\frac1{m_r}
\left[
\sigma_{Ni}\sigma_{ej}\{p_i,A_j\}
-
\sigma_{ei}\sigma_{Nj}\{p_i,A_j\}
\right].
\label{eq:O1CK}
\end{equation}
Since spin matrices belonging to different particles commute,
\begin{equation}
Y-X
=
\frac1{m_r}
\left(
\sigma_{Ni}\sigma_{ej}
-
\sigma_{ei}\sigma_{Nj}
\right)
\{p_i,A_j\}.
\label{eq:O2CK}
\end{equation}
Using
\begin{equation}
\{p_i,A_j\}
=
2A_jp_i-i\partial_iA_j,
\label{eq:anticommCK}
\end{equation}
the derivative is
\begin{equation}
\partial_iA_j
=
\frac{F_0(r)}{r}\delta_{ij}
+
\left[
F_0'(r)-\frac{F_0(r)}{r}
\right]\hat r_i\hat r_j .
\label{eq:derivACK}
\end{equation}
This tensor is symmetric under $i\leftrightarrow j$, whereas
\begin{equation}
C_{ij}
\equiv
\sigma_{Ni}\sigma_{ej}
-
\sigma_{ei}\sigma_{Nj}
\end{equation}
is antisymmetric. Consequently,
\begin{equation}
C_{ij}\partial_iA_j=0.
\label{eq:derivcancelCK}
\end{equation}
Thus all terms in which the momentum differentiates the radial
function cancel, and Eq.~\eqref{eq:O2CK} reduces to
\begin{equation}
Y-X
=
\frac{2F_0(r)}{m_r}
\left(
\sigma_{Ni}\sigma_{ej}
-
\sigma_{ei}\sigma_{Nj}
\right)
\hat r_j p_i.
\label{eq:O3CK}
\end{equation}
Using
\begin{equation}
\sigma_{Ni}\sigma_{ej}-\sigma_{ei}\sigma_{Nj}
=
-\epsilon_{ijk}(\bsig_e\times\bsig_N)_k,
\qquad
\rhat\times\bp
=
\frac{\br\times\bp}{r}
=
\frac{\bL}{r},
\end{equation}
we obtain
\begin{align}
\left(
\sigma_{Ni}\sigma_{ej}
-
\sigma_{ei}\sigma_{Nj}
\right)
\hat r_jp_i
&=
-\epsilon_{ijk}
(\bsig_e\times\bsig_N)_k\hat r_jp_i
\nonumber\\
&=
(\bsig_e\times\bsig_N)
\cdot\frac{\bL}{r}.
\label{eq:crossCK}
\end{align}
Therefore,
\begin{equation}
Y-X
=
\frac{2F_0(r)}{m_r r}
(\bsig_e\times\bsig_N)\cdot\bL
=
\frac{1}{2\pi m_r}
\frac{(\bsig_e\times\bsig_N)\cdot\bL}{r^3}.
\label{eq:TOPEantiCK}
\end{equation}

Substituting Eqs.~\eqref{eq:TOPEsym2} and
\eqref{eq:TOPEantiCK} into Eq.~\eqref{eq:CK21} gives
\begin{equation}
\begin{aligned}
V
=\frac{1}{4m_p m_r}\Bigl\{-(\beta_{eN}+\beta_{Ne})\,2\pi m_r\,(X+Y)\\
-(\beta_{eN}-\beta_{Ne})\,4\pi m_r\,\frac{F_0}{m_rr}(\bsig_1\times\bsig_2)\cdot\bL\Bigr\}
\end{aligned}
\end{equation}
\begin{equation}
\begin{aligned}
V=-\frac{\pi(\beta_{eN}+\beta_{Ne})}{2m_p}\,(X+Y)\\
-\frac{\pi(\beta_{eN}-\beta_{Ne})}{m_p}\,\frac{F_0}{m_r r}\,(\bsig_1\times\bsig_2)\cdot\bL.
\label{eq:CKinDM}
\end{aligned}
\end{equation}
Using Eq.~\eqref{eq:TOPEantiCK}, we can write our axial-pseudotensor TOPE equation Eq.~\eqref{eq:TOPEdecomp} as 
\begin{equation}
    V_{A \widetilde{T}}=\frac{A+B}{2}\,(X+Y)+(B-A)\,\frac{F_0}{m_r r}\,(\bsig_1\times\bsig_2)\cdot\bL.
\label{eq:ours}
\end{equation}

Comparing Eq.~\eqref{eq:ours} with Eq.~\eqref{eq:CKinDM}, we get the relation 
\begin{equation}
\frac{A+B}{2}=-\frac{\pi(\beta_{eN}+\beta_{Ne})}{2m_p},
\qquad
B-A=-\frac{\pi(\beta_{eN}-\beta_{Ne})}{m_p}.
\label{eq:twoconditions}
\end{equation}
Solving this, we obtain 
\begin{equation}
B=g_{\widetilde{T}}^eg_A^{N}=-\frac{\pi}{m_p}\,\beta_{eN},
\qquad
A=g_A^{e}g_{\widetilde{T}}^{N}=-\frac{\pi}{m_p}\,\beta_{Ne}.
\label{eq:dictionary}
\end{equation}

 $\beta_{eN}$ corresponds to the derivative
vertex on the electron, whereas $\beta_{Ne}$ corresponds to the
derivative vertex on the nucleon.

Therefore we obtain the limits on axial--vector-pseudotensor coupling constants with the relation 

\begin{equation}
|g_{\widetilde T}^{e}g_A^N|
=
\frac{\pi}{m_p}|\beta_{eN}|,
\qquad
|g_A^e g_{\widetilde T}^{N}|
=
\frac{\pi}{m_p}|\beta_{Ne}|
\label{eq:TOPEbetaabs}
\end{equation}
Thus Eq.~\eqref{eq:VATpt} reproduces the complete
Conti--Khriplovich electron--nucleon TOPE interaction in the
long-range limit and extends it to finite mediator mass.

\subsection{Tensor--tensor, tensor--pseudotensor, and pseudotensor--pseudotensor terms}

The two diagonal dipole combinations are compactly
\begin{align}
V_{TT}(\br)&=-4g_T^1g_T^2 Q_{12}(\br),
\label{eq:VTTspin1}\\
V_{\widetilde T\widetilde T}(\br)&=
4g_{\widetilde T}^1g_{\widetilde T}^2
\sigma_{1i}\sigma_{2j}H_{ij}(\br).
\label{eq:VPTPTspin1}
\end{align}
Both are $P,T$ even.  The mixed pseudotensor--tensor potential can be written
\begin{align}
V_{\widetilde T T}(\br)={}&
-\frac{2g_{\widetilde T}^1g_T^2}{m_2}\bsig_1\cdot\boldsymbol\nabla D_M
\nonumber\\
&+2g_{\widetilde T}^1g_T^2\,\sigma_{1i}
\left\{H_{ij},(\bsig_2\times\bv)_j\right\}
\nonumber\\
&+\frac{2g_T^1g_{\widetilde T}^2}{m_1}\bsig_2\cdot\boldsymbol\nabla D_M
\nonumber\\
&-2g_T^1g_{\widetilde T}^2\,\sigma_{2i}
\left\{H_{ij},(\bsig_1\times\bv)_j\right\}.
\label{eq:VTPTspin1}
\end{align}
This interaction is $P,T$ odd and corresponds to the momentum-suppressed $q^2V_{9,10}$ and $q^2V_{14}\pm V_{15}$ structures.

\section{Massive spin-2 mediator}
\label{sec:spin2}

A spin-2 particle is described by a symmetric rank-two field $h_{\mu\nu}=h_{\nu\mu}$.  It therefore cannot couple linearly to the antisymmetric current $\bar\psi\sigma^{\mu\nu}\psi$ without additional derivatives or fields.  The natural lowest-derivative fermion current is instead the symmetric energy--momentum tensor.  To explore parity violation we also introduce a symmetric axial-tensor current.

\subsection{Relativistic interaction and currents}

For fermion species $a$ define
\begin{align}
\Theta_a^{\mu\nu}={}&\frac{i}{4}\bar\psi_a
\left(\gamma^\mu\overleftrightarrow\partial^{\nu}
+\gamma^\nu\overleftrightarrow\partial^{\mu}\right)\psi_a
-\eta^{\mu\nu}\Lag_a,
\label{eq:ThetaDef}\\
\widetilde\Theta_a^{\mu\nu}={}&\frac{i}{4}\bar\psi_a
\left(\gamma^\mu\gamma_5\overleftrightarrow\partial^{\nu}
+\gamma^\nu\gamma_5\overleftrightarrow\partial^{\mu}\right)\psi_a.
\label{eq:Theta5Def}
\end{align}
We consider
\begin{equation}
\Lag_{2}=-h_{\mu\nu}\sum_a
\left(g_{2a}\Theta_a^{\mu\nu}
+\widetilde g_{2a}\widetilde\Theta_a^{\mu\nu}\right),
\label{eq:Lspin2}
\end{equation}
where $g_{2a}$ and $\widetilde g_{2a}$ have mass dimension $-1$.  For a graviton-like interaction $g_{2a}$ is usually written as a dimensionless coefficient divided by a common scale~\cite{HanLykkenZhang,Kang2021}.

Between on-shell spinors, with $K_a=p'_a+p_a$,
\begin{align}
T_a^{\mu\nu}&=\frac14\bar u'_a
\left(\gamma^\mu K_a^\nu+\gamma^\nu K_a^\mu\right)u_a,
\label{eq:Tmatrix}\\
\widetilde T_a^{\mu\nu}&=\frac14\bar u'_a
\left(\gamma^\mu\gamma_5 K_a^\nu
+\gamma^\nu\gamma_5 K_a^\mu\right)u_a.
\label{eq:T5matrix}
\end{align}
For equal-mass free on-shell external fermions, the matrix element of the Dirac Lagrangian vanishes by the Dirac equation,
$\langle p'_a|\mathcal{L}_a|p_a\rangle=0$,
so the $-\eta^{\mu\nu}\mathcal{L}_a$ term in
Eq.~\eqref{eq:ThetaDef} does not contribute to
$T_a^{\mu\nu}$.

The ordinary current is conserved,
\begin{equation}
q_\mu T_a^{\mu\nu}=0,
\label{eq:Tconserved}
\end{equation}
whereas the axial-tensor current is traceless but nonconserved for $m_a\ne0$:
\begin{align}
\widetilde T_{a\,\mu}^{\ \ \mu}&=0,
\label{eq:T5trace}\\
q_\mu\widetilde T_1^{\mu\nu}&=
\frac{m_1}{2}K_1^\nu\bar u'_1\gamma_5u_1,
\label{eq:T5div1}\\
q_\mu\widetilde T_2^{\mu\nu}&=
-\frac{m_2}{2}K_2^\nu\bar u'_2\gamma_5u_2,
\label{eq:T5div2}\\
q_\mu q_\nu\widetilde T_a^{\mu\nu}&=0.
\label{eq:T5double}
\end{align}
The last identity follows from $q\cdot K_a=0$ in elastic scattering.  

\subsection{Fierz--Pauli propagator and longitudinal helicities}

The massive spin-2 propagator can be written~\cite{Koenigstein,Hinterbichler}
\begin{equation}
D_{\mu\nu,\alpha\beta}(q)=
\frac{i}{q^2-M^2+i0}\,
\mathcal P_{\mu\nu,\alpha\beta}(q),
\label{eq:spin2prop}
\end{equation}
with
\begin{align}
\mathcal P_{\mu\nu,\alpha\beta}={}&\frac12
\left(\theta_{\mu\alpha}\theta_{\nu\beta}
+\theta_{\mu\beta}\theta_{\nu\alpha}\right)
-\frac13\theta_{\mu\nu}\theta_{\alpha\beta},
\label{eq:spin2projector}\\
\theta_{\mu\nu}={}&\eta_{\mu\nu}-\frac{q_\mu q_\nu}{M^2}.
\label{eq:theta}
\end{align}
For two symmetric sources $J^{\mu\nu}$ and $K^{\mu\nu}$,
\begin{align}
J\mathcal P K={}&J^{\mu\nu}K_{\mu\nu}-\frac13JK
\nonumber\\
&-\frac{2}{M^2}(q_\mu J^{\mu\nu})(q^\alpha K_{\alpha\nu})
\nonumber\\
&+\frac{1}{3M^2}\left[J(qKq)+K(qJq)\right]
\nonumber\\
&+\frac{2}{3M^4}(qJq)(qKq),
\label{eq:saturated}
\end{align}
where $J=J^\mu{}_{\mu}$ and $qJq=q_\mu J^{\mu\nu}q_\nu$.  Equation~\eqref{eq:saturated} shows explicitly why all longitudinal pieces vanish for two conserved energy--momentum currents.  They also vanish in the mixed $T\widetilde T$ amplitude because $T$ is conserved and $\widetilde T$ is traceless with $q\widetilde Tq=0$.  In contrast, for two axial-tensor currents,
\begin{equation}
\widetilde T_1\mathcal P\widetilde T_2=
\widetilde T_1^{\mu\nu}\widetilde T_{2\mu\nu}
-\frac{2}{M^2}
(q_\mu\widetilde T_1^{\mu\nu})
(q^\alpha\widetilde T_{2\alpha\nu}),
\label{eq:T5PT5}
\end{equation}
so the longitudinal helicities contribute for massive fermions. Since $\widetilde T^{\mu\nu}$ is not conserved and generates a term proportional to $1/M^2$ in the $\widetilde T\widetilde T$ channel, the resulting $1/M^2$ enhancement is a signal of sensitivity to its ultraviolet completion. The axial-tensor spin-2 interaction should therefore be understood as an effective description with a finite cutoff and not extrapolated to arbitrarily small mediator mass.

The master nonrelativistic potential is
\begin{equation}
\Vt_2(\bq)=-\frac{1}{4m_1m_2}
\frac{J_1^{\mu\nu}\mathcal P_{\mu\nu,\alpha\beta}J_2^{\alpha\beta}}
{\bq^2+M^2},
\label{eq:spin2master}
\end{equation}
with $J_a^{\mu\nu}=g_{2a}T_a^{\mu\nu}+\widetilde g_{2a}\widetilde T_a^{\mu\nu}$.

\subsection{Nonrelativistic currents}

To the leading orders needed here,
\begin{align}
T_1^{00}&\simeq2m_1^2,
&T_1^{0i}&\simeq2m_1p_1^i-\frac{i m_1}{2}(\bq\times\bsig_1)^i,
\label{eq:TNR1}\\
T_2^{00}&\simeq2m_2^2,
&T_2^{0i}&\simeq2m_2p_2^i+\frac{i m_2}{2}(\bq\times\bsig_2)^i,
\label{eq:TNR2}
\end{align}
and
\begin{align}
\widetilde T_a^{00}&\simeq2m_a\bsig_a\cdot\bp_a,
\label{eq:T5NR00}\\
\widetilde T_a^{0i}&\simeq m_a^2\sigma_a^i,
\label{eq:T5NR0i}\\
\widetilde T_a^{ij}&\simeq m_a
\left(p_a^i\sigma_a^j+p_a^j\sigma_a^i\right).
\label{eq:T5NRij}
\end{align}
The displayed currents are sufficient for the static and mixed terms below.  In deriving the spin--orbit term in Eq.~\eqref{eq:spin2SOq} we additionally retain the subleading spin-dependent pieces of the ordinary energy--momentum tensor that are quadratic in nonrelativistic momenta; still higher terms are otherwise omitted.

\subsection{Tensor--tensor spin-2 exchange}

For two conserved currents, the leading central and static spin-dependent potential is
\begin{align}
\Vt_{22}(\bq)={}&
-\frac{2g_{21}g_{22}m_1m_2}{3(\bq^2+M^2)}
\nonumber\\
&+\frac{g_{21}g_{22}}{8(\bq^2+M^2)}
\bigl[\bq^2\bsig_1\cdot\bsig_2
\nonumber\\
&\qquad\qquad-(\bsig_1\cdot\bq)(\bsig_2\cdot\bq)\bigr].
\label{eq:spin2TTq}
\end{align}
Therefore
\begin{align}
V_{22}(\br)={}&-\frac23g_{21}g_{22}m_1m_2Y_M(r)
\nonumber\\
&+\frac18g_{21}g_{22}Q_{12}(\br)+O(v^2).
\label{eq:spin2TTr}
\end{align}
At the same quadratic order in nonrelativistic momenta there is also a spin--orbit term.
Keeping the terms linear in $\bP$ and $\bq$ gives
\begin{align}
\widetilde V_{22}^{\rm SO}(\bq)={}&
\frac{i g_{21}g_{22}}{\bq^2+M^2}
\left(a_1\bsig_1+a_2\bsig_2\right)\cdot(\bP\times\bq),
\label{eq:spin2SOq}\\
 a_1={}&\frac12+\frac{m_2}{3m_1},\qquad
 a_2=\frac12+\frac{m_1}{3m_2}.
\label{eq:spin2acoef}
\end{align}
After the Hermitian replacement in Eq.~\eqref{eq:symm},
\begin{equation}
 V_{22}^{\rm SO}(\br)=-\frac{g_{21}g_{22}}{2}
 \left\{\left(a_1\bsig_1+a_2\bsig_2\right)\cdot
 (\bP\times\rhat),F_M(r)\right\}.
\label{eq:spin2SOr}
\end{equation}
This is the $V_{4,5}$ sector.  For $m_1=m_2$, $a_1=a_2=5/6$ and only the symmetric $V_{4+5}$ combination remains.
The first term in Eq.~\eqref{eq:spin2TTr} is the massive-spin-2 Yukawa force; the second contains the $V_2+V_3$ spin--spin and tensor structures, including their contact completion.  All terms in Eqs.~\eqref{eq:spin2TTr} and \eqref{eq:spin2SOr} conserve $P$ and $T$.  At still higher order the conventional energy--momentum coupling also produces the $q^2V_8$ structure; this is consistent with the complete fermionic spin-2 expansion of Ref.~\cite{Kang2021}.

\subsection{Tensor--axial-tensor spin-2 exchange}

For one ordinary and one axial-tensor vertex we find
\begin{align}
\Vt_{2\widetilde2}(\bq)={}&
\frac{g_{21}\widetilde g_{22}m_1m_2}{\bq^2+M^2}
\bsig_2\cdot\bv
\nonumber\\
&-\frac{i g_{21}\widetilde g_{22}m_2}{4(\bq^2+M^2)}
(\bsig_1\times\bsig_2)\cdot\bq
\nonumber\\
&-\frac{\widetilde g_{21}g_{22}m_1m_2}{\bq^2+M^2}
\bsig_1\cdot\bv
\nonumber\\
&-\frac{i\widetilde g_{21}g_{22}m_1}{4(\bq^2+M^2)}
(\bsig_1\times\bsig_2)\cdot\bq.
\label{eq:spin2T5q}
\end{align}
In coordinate space,
\begin{align}
V_{2\widetilde2}(\br)={}&
\frac{m_1m_2g_{21}\widetilde g_{22}}{2}
\{\bsig_2\cdot\bv,Y_M\}
\nonumber\\
&-\frac{m_1m_2\widetilde g_{21}g_{22}}{2}
\{\bsig_1\cdot\bv,Y_M\}
\nonumber\\
&+\frac{m_2g_{21}\widetilde g_{22}
+m_1\widetilde g_{21}g_{22}}{4}
(\bsig_1\times\bsig_2)\cdot\rhat\,F_M.
\label{eq:spin2T5r}
\end{align}
The first line has the $V_{12,13}$ spin--velocity structure and the second the $V_{11}$ structure.  Both are $P$ odd and $T$ even.  Consequently the minimal one-derivative tensor--axial-tensor coupling of a massive spin-2 boson violates parity but not time reversal.

\subsection{Axial-tensor--axial-tensor spin-2 exchange}

Using Eq.~\eqref{eq:T5PT5}, the leading result is
\begin{align}
\Vt_{\widetilde2\widetilde2}(\bq)={}&
\frac{\widetilde g_{21}\widetilde g_{22}m_1m_2}
{2(\bq^2+M^2)}
\nonumber\\
&\times\left[\bsig_1\cdot\bsig_2
+\frac{(\bsig_1\cdot\bq)(\bsig_2\cdot\bq)}{M^2}\right].
\label{eq:spin255q}
\end{align}
Its coordinate-space form is
\begin{align}
V_{\widetilde2\widetilde2}(\br)={}&
\frac{\widetilde g_{21}\widetilde g_{22}m_1m_2}{2}
\bigl[(\bsig_1\cdot\bsig_2)Y_M(r)
\nonumber\\
&+M^{-2}\sigma_{1i}\sigma_{2j}H_{ij}(\br)\bigr]+O(v^2).
\label{eq:spin255r}
\end{align}
This potential is $P,T$ even.  The explicit $1/M^2$ contribution is due to the longitudinal helicities and reflects the nonconservation of $\widetilde\Theta^{\mu\nu}$ for a massive fermion.

\section{Relation to the sixteen rotationally invariant potentials}
\label{sec:symmetry}

For comparison with experimental searches we now make the relation of the
relativistic interactions above to the standard sixteen-operator basis of
Dobrescu and Mocioiu explicit~\cite{DobrescuMocioiu,Cong2025}.  In the center-of-mass frame we use the momentum $\bP$ defined in
Eq.~\eqref{eq:Preduced}, with $\bq=\bp'_1-\bp_1$ and
$\bq\cdot\bP=0$, and introduce an arbitrary reference mass $m_0$ only to
make the operators dimensionless.  A convenient Hermitian basis is
\begin{align}
{\cal O}_1&=1,
\qquad
{\cal O}_2=\bsig_1\cdot\bsig_2,
\nonumber\\
{\cal O}_3&=\frac{(\bsig_1\cdot\bq)(\bsig_2\cdot\bq)}{m_0^2},
\nonumber\\
{\cal O}_{4,5}&=\frac{i}{2m_0^2}(\bsig_1\pm\bsig_2)\cdot(\bP\times\bq),
\nonumber\\
{\cal O}_{6,7}&=\frac{i}{2m_0^2}
 \bigl[(\bsig_1\cdot\bP)(\bsig_2\cdot\bq)
\nonumber\\
&\quad\pm(\bsig_1\cdot\bq)(\bsig_2\cdot\bP)\bigr],
\nonumber\\
{\cal O}_8&=\frac{(\bsig_1\cdot\bP)(\bsig_2\cdot\bP)}{m_0^2},
\nonumber\\
{\cal O}_{9,10}&=\frac{i}{2m_0}(\bsig_1\pm\bsig_2)\cdot\bq,
\nonumber\\
{\cal O}_{11}&=\frac{i}{m_0}(\bsig_1\times\bsig_2)\cdot\bq,
\nonumber\\
{\cal O}_{12,13}&=\frac{1}{2m_0}(\bsig_1\pm\bsig_2)\cdot\bP,
\qquad
{\cal O}_{14}=\frac{(\bsig_1\times\bsig_2)\cdot\bP}{m_0},
\nonumber\\
{\cal O}_{15}&=\frac{1}{2m_0^3}\Bigl\{
 [\bsig_1\cdot(\bP\times\bq)](\bsig_2\cdot\bq)
\nonumber\\
&\quad+(\bsig_1\cdot\bq)[\bsig_2\cdot(\bP\times\bq)]\Bigr\},
\nonumber\\
{\cal O}_{16}&=\frac{i}{2m_0^3}\Bigl\{
 [\bsig_1\cdot(\bP\times\bq)](\bsig_2\cdot\bP)
\nonumber\\
&\quad+(\bsig_1\cdot\bP)[\bsig_2\cdot(\bP\times\bq)]\Bigr\}.
\label{eq:DMbasis}
\end{align}
Scalar functions of $\bq^2$ and $\bP^2$ multiply these operators, so a factor
$\bq^2$ does not define a new rotational structure.  We nevertheless keep an
explicit notation such as $q^2V_i$ below because it is physically important:
it changes the radial dependence and, in particular, may turn a long-range
potential into a contact-dominated one.

The discrete symmetries of the basis are
\begin{equation}
\begin{array}{c|c|c}
 & P\text{-even} & P\text{-odd}\\ \hline
T\text{-even} & V_{1,2,3,4,5,8} & V_{11,12,13,16}\\
T\text{-odd}  & V_{6,7} & V_{9,10,14,15}
\end{array}.
\label{eq:PTbasis}
\end{equation}
In particular, $V_{6,7}$ are the only $P$-even, $T$-odd structures in this
basis.

Table~\ref{tab:mapping} gives the mapping of the interactions derived in this
work.  For spin 1, the tensor and pseudotensor labels refer to the dipole
vertices $q_\mu\sigma^{\mu\nu}$ and $q_\mu\sigma^{\mu\nu}\gamma_5$.
For spin 2, ``tensor'' denotes the symmetric energy--momentum current and
``axial tensor'' its symmetric $\gamma_5$ counterpart; these are different
Lorentz representations from the spin-1 dipole operators.

\begin{table*}[t]
\caption{Mapping of the new spin-1 and spin-2 interactions onto the sixteen
rotationally invariant potentials of Ref.~\cite{DobrescuMocioiu}.  An explicit
$q^2$ indicates a momentum-suppressed realization of the same $V_i$ operator.
The entries refer to the orders retained in the present calculation.}
\label{tab:mapping}
\centering
\small
\begin{tabular}{lllll}
\toprule
Mediator & Relativistic coupling product & $V_i$ structures & $P$ & $T$\\
\midrule
spin 1 & $g_Vg_T$ & $q^2V_1$, $V_{4,5}$, $V_2+V_3$ & even & even\\
spin 1 & $g_Vg_{\widetilde T}$ & $V_{9,10}$, $q^2V_{14}\pm V_{15}$ & odd & odd\\
spin 1 & $g_Ag_T$ & $q^2V_{12,13}$, $V_{11}$ & odd & even\\
spin 1 & $g_Ag_{\widetilde T}$ & $V_{6,7}$ & even & odd\\
spin 1 & $g_Tg_T$ & $V_2+V_3$ & even & even\\
spin 1 & $g_{\widetilde T}g_T$ & $q^2V_{9,10}$, $q^2V_{14}\pm V_{15}$ & odd & odd\\
spin 1 & $g_{\widetilde T}g_{\widetilde T}$ & $V_3$ & even & even\\
\midrule
spin 2 & $g_2g_2$ & $V_1$, $V_2+V_3$, $V_{4,5}$; $q^2V_8$ at higher order & even & even\\
spin 2 & $g_2\widetilde g_2$ & $V_{12,13}$, $V_{11}$ & odd & even\\
spin 2 & $\widetilde g_2\widetilde g_2$ & $V_2+V_3$ & even & even\\
\bottomrule
\end{tabular}
\end{table*}

\begin{figure*}[t]
\centering
\includegraphics[width=\textwidth]{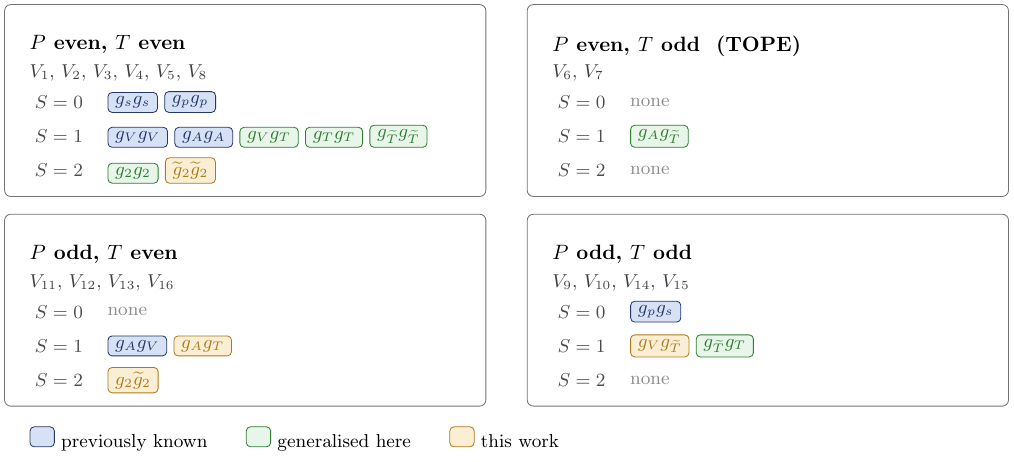}
\caption{Mapping of single-boson exchange onto the sixteen rotationally
invariant potentials, organized by the discrete symmetries of
Eq.~\eqref{eq:PTbasis}. Each panel names the $V_i$ belonging to that $P,T$
class and lists the exchange channels that populate it for a spin-0 ($S=0$), spin-1 ($S=1$) and spin-2
($S=2$) mediator. The $S=0$ entries are the scalar/pseudoscalar potentials
of Refs.~\cite{MoodyWilczek,Fadeev2019}; $g_2$ and $\widetilde g_2$ denote the
energy--momentum and axial-tensor spin-2 couplings of
Eq.~\eqref{eq:Lspin2}. Blue entries denote previously known interactions,
green entries denote interactions generalized here, and orange entries
denote the remaining channels derived in this work. In particular, the
spin-1 vector--tensor $g_Vg_T$ interaction is generalized to arbitrary fermion masses and independent couplings at the two vertices, and $g_Ag_{\widetilde T}$ TOPE interaction is generalized to finite mediator mass, while the conventional spin-2 $g_2g_2$ interaction is rederived and generalized in the present framework. The parity-even, time-reversal-odd sector is populated only by the spin-1 $g_Ag_{\widetilde T}$ cross term, while the minimal spin-2 interactions populate only the $T$-even sectors.}
\label{f:mapping}
\end{figure*}

The same information is displayed in Fig.~\ref{f:mapping}, organized by the discrete symmetries rather than by the relativistic coupling.
Several observations follow immediately. First, the finite-range  spin-1 cross terms considered here fill precisely the gap left by the usual scalar/pseudoscalar and vector/axial-vector exchanges: $g_Ag_{\widetilde T}$ produces $V_{6,7}$ at
tree level.  Second, the spin-2 basis considered here produces only the $T$-even sectors of Eq.~\eqref{eq:PTbasis}; the conventional $g_2g_2$ interaction generates $V_{4,5}$ already at the same quadratic momentum order as its static spin--spin term, while $q^2V_8$ starts at the next order.  Hence no choice of real diagonal
$g_2$ and $\widetilde g_2$ generates either the TOPE $V_{6,7}$ structures or
the $P,T$-odd $V_{9,10,14,15}$ structures at this derivative order.  Third, several spin-1 and spin-2 amplitudes share an identical $V_i$ component and radial function.  Those identities permit direct coefficient reinterpretations of existing limits, but only when the experimental limit was obtained from the shared component.  A complete relativistic coupling may generate additional operators, so equality of one $V_i$ component must not be confused with equality of the full potentials.

For later use, two such component identities are particularly simple.  From
Eqs.~\eqref{eq:VTTspin1} and \eqref{eq:spin2TTr},
\begin{align}
 \left.V_{22}\right|_{Q_{12}}&=\frac{g_{21}g_{22}}{8}Q_{12},
\label{eq:S2S1Q}\\
 \left.V_{TT}^{(1)}\right|_{Q_{12}}&=-4g_T^1g_T^2Q_{12},
\label{eq:S1TTQ}
\end{align}
so equal $V_2+V_3$ components correspond to
\begin{equation}
 g_{21}g_{22}=-32g_T^1g_T^2.
\label{eq:mapS2TT}
\end{equation}
It is also instructive to compare ordinary spin-1 vector exchange with the
spin-2 energy--momentum coupling.  To the same static order,
\begin{equation}
 V_{VV}^{(1)}=g_V^1g_V^2Y_M-\frac{g_V^1g_V^2}{4m_1m_2}Q_{12}+\cdots .
\label{eq:S1VVcompare}
\end{equation}
The $V_1$ terms alone would be equal for
\begin{equation}
 g_{21}g_{22}=-\frac{3}{2m_1m_2}g_V^1g_V^2,
\label{eq:S2S1V1}
\end{equation}
whereas equality of the $V_2+V_3$ components requires
\begin{equation}
 g_{21}g_{22}=-\frac{2}{m_1m_2}g_V^1g_V^2.
\label{eq:S2S1V23}
\end{equation}
The two rescalings are different, so a spin-1 vector potential and a spin-2
energy--momentum potential are not globally proportional even though they
share the same $V_1$ and $V_2+V_3$ operator structures.

Likewise the common $V_{11}$ component of Eqs.~\eqref{eq:VATcompact} and
\eqref{eq:spin2T5r} gives, for the ordering shown,
\begin{align}
 -2g_A^1g_T^2&=\frac{m_2}{4}g_{21}\widetilde g_{22},
\nonumber\\
 \text{i.e.}\qquad
 g_{21}\widetilde g_{22}&=-\frac{8}{m_2}g_A^1g_T^2.
\label{eq:mapS2V11}
\end{align}
The qualifications are important: $V_{22}$ also contains the central $V_1$
term, while $V_{2\widetilde2}$ contains an unsuppressed $V_{12,13}$ term.
Equations~\eqref{eq:mapS2TT} and \eqref{eq:mapS2V11} are therefore
operator-by-operator translations, not universal rescalings of the full
spin-2 models.

A third useful component identity follows from the $H_{ij}$, or $V_3$,
part of Eqs.~\eqref{eq:VPTPTspin1} and \eqref{eq:spin255r}.  At fixed
mediator mass,
\begin{align}
 4g_{\widetilde T}^1g_{\widetilde T}^2
 &=\frac{m_1m_2}{2M^2}\widetilde g_{21}\widetilde g_{22},
\nonumber\\
 \text{i.e.}\qquad
 \widetilde g_{21}\widetilde g_{22}
 &=\frac{8M^2}{m_1m_2}g_{\widetilde T}^1g_{\widetilde T}^2.
\label{eq:mapS2PTPT}
\end{align}
This maps the identical $V_3$ radial function only.  The spin-2
$\widetilde g_2\widetilde g_2$ interaction additionally contains the
isotropic $V_2$ term.  Alternatively, its complete leading radial bracket is
identical to the ordinary spin-1 axial--axial Proca potential, for which
\begin{equation}
 \widetilde g_{21}\widetilde g_{22}
 =-\frac{2g_A^1g_A^2}{m_1m_2}
\label{eq:mapS2AAexact}
\end{equation}
(up to the overall interaction-sign convention).  Thus axial--axial limits
may be converted directly when the published analysis uses the full
$V_2+V_3$ response in the same normalization.

\section{Heavy-mediator contact limits}
\label{sec:contact}

For $M^2\gg\bq^2$,
\begin{align}
\frac{1}{\bq^2+M^2}&=\frac1{M^2}-\frac{\bq^2}{M^4}+\cdots,
\label{eq:contactexp}\\
Y_M(r)&\longrightarrow\frac{\delta^3(\br)}{M^2}.
\label{eq:contactY}
\end{align}
For spin 2 the tensor--tensor interaction becomes
\begin{align}
V_{22}^{\rm contact}={}&-\frac{2g_{21}g_{22}m_1m_2}{3M^2}\delta^3(\br)
\nonumber\\
&-\frac{g_{21}g_{22}}{8M^2}
(\bsig_1\cdot\bsig_2)\nabla^2\delta^3(\br)
\nonumber\\
&+\frac{g_{21}g_{22}}{8M^2}
(\bsig_1\cdot\boldsymbol\nabla)
(\bsig_2\cdot\boldsymbol\nabla)\delta^3(\br)
+O(M^{-4}).
\label{eq:spin2TTcontact}
\end{align}
The parity-odd mixed interaction gives
\begin{align}
V_{2\widetilde2}^{\rm contact}={}&
\frac{m_1m_2g_{21}\widetilde g_{22}}{2M^2}
\{\bsig_2\cdot\bv,\delta^3(\br)\}
\nonumber\\
&-\frac{m_1m_2\widetilde g_{21}g_{22}}{2M^2}
\{\bsig_1\cdot\bv,\delta^3(\br)\}
\nonumber\\
&-\frac{m_2g_{21}\widetilde g_{22}}{4M^2}
(\bsig_1\times\bsig_2)\cdot\boldsymbol\nabla\delta^3(\br)
\nonumber\\
&-\frac{m_1\widetilde g_{21}g_{22}}{4M^2}
(\bsig_1\times\bsig_2)\cdot\boldsymbol\nabla\delta^3(\br)
+O(M^{-4}).
\label{eq:spin2T5contact}
\end{align}
The leading axial-tensor--axial-tensor result is
\begin{align}
V_{\widetilde2\widetilde2}^{\rm contact}={}&
\frac{\widetilde g_{21}\widetilde g_{22}m_1m_2}{2M^2}
(\bsig_1\cdot\bsig_2)\delta^3(\br)
\nonumber\\
&-\frac{\widetilde g_{21}\widetilde g_{22}m_1m_2}{2M^4}
(\bsig_1\cdot\boldsymbol\nabla)(\bsig_2\cdot\boldsymbol\nabla)\delta^3(\br)+\cdots.
\label{eq:spin255contact}
\end{align}
These spin-2 contact operators are not the conventional atomic-EDM tensor--pseudotensor interaction
$(\bar e\sigma_{\mu\nu}e)(\bar N i\sigma^{\mu\nu}\gamma_5N)$, whose Lorentz indices are antisymmetric~\cite{GingesFlambaum}.  A pure symmetric spin-2 field cannot couple linearly to that antisymmetric current.

\section{Massless limit and the graviton}
\label{sec:massless}

The $M\to0$ limit of a massive Fierz--Pauli propagator is not identical to the propagator of a genuinely massless spin-2 gauge field.  For conserved sources the massive exchange contains
\begin{equation}
T_1^{\mu\nu}T_{2\mu\nu}-\frac13T_1T_2,
\end{equation}
whereas the massless graviton contains
\begin{equation}
T_1^{\mu\nu}T_{2\mu\nu}-\frac12T_1T_2.
\label{eq:vDVZ}
\end{equation}
This is the linear van Dam--Veltman--Zakharov discontinuity; nonlinear screening in consistent massive-gravity theories changes the physical interpretation at sufficiently short distances~\cite{Hinterbichler}.

For a genuine massless graviton, gauge invariance requires coupling to a conserved symmetric source.  With
\begin{equation}
\Lag_{\rm grav}=-\frac{\kappa}{2}h_{\mu\nu}T^{\mu\nu},
\qquad \kappa^2=32\pi G,
\end{equation}
the leading nonrelativistic potential is the Newton interaction
\begin{equation}
V_N(r)=-\frac{Gm_1m_2}{r},
\end{equation}
which is $P,T$ even.  The axial-tensor current of Eq.~\eqref{eq:Theta5Def} is not conserved for massive electrons, nucleons, or quarks and therefore cannot be introduced as an independent minimal source for an ordinary massless graviton.  The same conservation requirement restricts the couplings available to strictly massless spin-1 fields~\cite{Dobrescu2005}.  In particular, the $1/M^2$ longitudinal term in Eq.~\eqref{eq:spin255r} has no smooth massless-graviton limit.

\section{Existing limits and their reinterpretation}
\label{sec:bounds}

\subsection{Connection with the conventions and exclusion plots of Ref.~\cite{Cong2025}}
\label{subsec:RMPdictionary}

Review of Modern Physics paper ~\cite{Cong2025} (abbreviated as RMP) is particularly useful for converting the results of
this paper into experimentally constrained quantities.  It presents both the
Dobrescu--Mocioiu (DM) convention, in which a potential is written as
$\mathcal V_i=f_i^{XY}V_i$, and the coordinate-space particle-exchange
convention of Fadeev \emph{et al.}, in which the coefficients are products of
physical fermion--boson couplings.  Appendix A of Ref.~\cite{Cong2025}, and in
particular its Table VI, gives the conversion between these conventions.  We
adopt its normalization in this subsection.  The labels $X$ and $Y$ specify
the ordering of the two fermions, and
\begin{align}
 m_r=\frac{m_Xm_Y}{m_X+m_Y},\qquad M=\lambda^{-1}
\label{eq:RMPconv}
\end{align}
in natural units.  Overall signs depend on the sign convention chosen for the
interaction Lagrangian; limits below therefore use absolute values.

In terms of the kernels in Eqs.~\eqref{eq:Y}--\eqref{eq:H}, the RMP/DM
potentials relevant to the interactions derived here may be written in the
compact forms collected in Table~\ref{tab:RMPpotentials}.  These are the
quantum, center-of-mass forms (including contact terms where appropriate).
For macroscopic nonoverlapping sources the delta functions are omitted and
$\bp_X=m_r\bv$ may be treated classically.  The $V_{6+7}$ line is written in
the original DM normalization because this term is not included in Table VI
of Ref.~\cite{Cong2025}; Ref.~\cite{Cong2025} nevertheless explicitly notes
that the extended spin-1 Lagrangian containing $g_Ag_{\widetilde T}$ produces
$V_{6,7}$.

\begin{table*}[t]
\caption{Relevant Dobrescu--Mocioiu/RMP potentials in natural units.  The
normalization follows Appendix A of Ref.~\cite{Cong2025}.  $H_{ij}$ is defined
in Eq.~\eqref{eq:H}.  The full $V_{14q+15}$ expression is Eq.~(A9) of
Ref.~\cite{Cong2025}; only its characteristic operator content is displayed
here.}
\label{tab:RMPpotentials}
\centering\scriptsize
\renewcommand{\arraystretch}{1.25}
\begin{tabular}{@{}p{0.105\textwidth}p{0.52\textwidth}p{0.27\textwidth}@{}}
\toprule
RMP/DM term & Potential $\mathcal V_i=f_i^{XY}V_i$ & Standard Fadeev/RMP realization used for limits \\
\midrule
$V_1$ & $f_1^{XY}Y_M(r)$ & $V_1|_{VV}=g_V^Xg_V^Y Y_M$ \\
$V_2$ & $f_2^{XY}(\bsig_X\!\cdot\!\bsig_Y)Y_M$ & $f_2^{XY}=-g_A^Xg_A^Y$ \\
$V_3$ & $\displaystyle {f_3^{XY}\over m_X^2}\,\sigma_{Xi}\sigma_{Yj}H_{ij}$ & $\displaystyle f_3^{XY}={m_X\over4m_Y}g_V^Xg_V^Y=-{m_X\over4m_Y}g_p^Xg_p^Y$; also $\displaystyle f_3^{XY}=-{m_X^2\over M^2}g_A^Xg_A^Y$ \\
$V_{4+5}$ & $\displaystyle -{f_{4+5}^{XY}\over4m_X^2}\{\bsig_X\!\cdot(\bp_X\!\times\!\rhat),F_M\}$ & $\displaystyle f_{4+5}^{XY}=-\left({1\over2}+{m_X\over m_Y}\right)g_V^Xg_V^Y$ (vector--vector) \\
$V_{6+7}$ & $\displaystyle -{f_{6+7}^{XY}\over2m_X}(\bsig_X\!\cdot\!\bv)(\bsig_Y\!\cdot\!\rhat)F_M$ & Generated by the $g_A^Xg_{\widetilde T}^Y$ cross term of the extended spin-1 Lagrangian \\
$V_8$ & $\displaystyle {f_8^{XY}\over4m_X^2}\{\bsig_X\!\cdot\!\bp_X,\{\bsig_Y\!\cdot\!\bp_X,Y_M\}\}$ & $\displaystyle f_8^{XY}=-{1\over2}\left(1+{2m_X\over m_Y}+{m_X^2\over m_Y^2}\right)g_A^Xg_A^Y$ \\
$V_{9+10}$ & $\displaystyle -{f_{9+10}^{XY}\over2m_X}(\bsig_X\!\cdot\!\rhat)F_M$ & $f_{9+10}^{XY}=g_p^Xg_s^Y$ \\
$V_{11}$ & $\displaystyle -{f_{11}^{XY}\over m_X}(\bsig_X\!\times\!\bsig_Y)\!\cdot\!\rhat\,F_M$ & $\displaystyle f_{11}^{XY}={1\over2}g_V^Xg_A^Y+{m_X\over2m_Y}g_A^Xg_V^Y$ \\
$V_{12+13}$ & $\displaystyle {f_{12+13}^{XY}\over4m_X}\{\bsig_X\!\cdot\!\bp_X,Y_M\}$ & $\displaystyle f_{12+13}^{XY}=2\left(1+{m_X\over m_Y}\right)g_A^Xg_V^Y$ \\
$V_{14q+15}$ & $f_{14q+15}^{XY}$ times the common $q^2V_{14}+V_{15}$ kernel, containing $(\bsig_X\!\times\!\bsig_Y)\!\cdot\!\bp_X$ and the associated tensor term & $\displaystyle f_{14q+15}^{XY}=-{1\over4}{m_X^2\over m_Y^2}g_p^Xg_s^Y$ for the pseudoscalar--scalar realization \\
\bottomrule
\end{tabular}
\end{table*}

For reference, the leading Fadeev/RMP potentials needed most often in the
translations can be written especially simply in the notation of this paper,
\begin{align}
V_{VV}&=g_V^Xg_V^Y Y_M-
 {g_V^Xg_V^Y\over4m_Xm_Y}Q_{XY}+\cdots,
\label{eq:RMPVVcompact}\\
V_{AA}&=-g_A^Xg_A^Y\left[(\bsig_X\!\cdot\!\bsig_Y)Y_M+
 {\sigma_{Xi}\sigma_{Yj}H_{ij}\over M^2}\right]+\cdots,
\label{eq:RMPAAcompact}\\
V_{AV}^{XY}&=g_A^Xg_V^Y(\bsig_X\!\cdot\!\bv)Y_M
-{g_A^Xg_V^Y\over2m_Y}(\bsig_X\!\times\!\bsig_Y)\!\cdot\!\rhat F_M+\cdots,
\label{eq:RMPAVcompact}\\
V_{ps}^{XY}&=-{g_p^Xg_s^Y\over2m_X}(\bsig_X\!\cdot\!\rhat)F_M+\cdots,
\label{eq:RMPpscompact}\\
V_{pp}&=-{g_p^Xg_p^Y\over4m_Xm_Y}\sigma_{Xi}\sigma_{Yj}H_{ij}.
\label{eq:RMPppcompact}
\end{align}
The ellipses denote the paired higher-order terms displayed explicitly in
Eqs.~(16)--(24) of Ref.~\cite{Cong2025}.

The most useful result for the present paper is the coefficient dictionary in
Table~\ref{tab:newRMPmatch}.  It permits a limit on either an RMP coefficient
$f_i^{XY}$ or a Fadeev-style coupling product to be converted directly into
a limit on a coupling product introduced here.  The table is intentionally
component-by-component: when two relativistic models generate the same
$V_i$ with different ratios to other $V_j$ terms, there is no universal
rescaling of the complete potentials.

\begin{table*}[t]
\caption{Component-by-component matching of the new spin-1 and spin-2
couplings to the RMP conventions.  Magnitudes are shown because the exclusion
limits are bounds on absolute values.  $a_X=1/2+m_Y/(3m_X)$.  ``Equivalent
RMP product'' means the physical coupling product that gives the same stated
$V_i$ component, not an equality of complete relativistic models.}
\label{tab:newRMPmatch}
\centering\scriptsize
\renewcommand{\arraystretch}{1.2}
\begin{tabular}{@{}p{0.09\textwidth}p{0.18\textwidth}p{0.105\textwidth}p{0.275\textwidth}p{0.27\textwidth}@{}}
\toprule
Mediator & New product $G$ & Shared term & DM/RMP coefficient & Equivalent RMP physical product \\
\midrule
spin 1 & $|g_T^Xg_V^Y|$ & $V_{4+5}$ & $\displaystyle |f_{4+5}^{XY}|={4m_X^2\over m_r}|G|$ & $\displaystyle |g_V^Xg_V^Y|_{4+5}={4m_X^2\over m_r(1/2+m_X/m_Y)}|G|$ \\
spin 1 & $|g_V^Xg_T^Y|$ & $V_3$ component & $\displaystyle |f_3^{XY}|={m_X}|G|$ & $|g_V^Xg_V^Y|_{3}=|g_p^Xg_p^Y|_3=4m_Y|G|$ \\
spin 1 & $|g_{\widetilde T}^Xg_V^Y|$ & $V_{9+10}$ & $|f_{9+10}^{XY}|=4m_X|G|$ & $|g_p^Xg_s^Y|=4m_X|G|$ \\
spin 1 & $|g_A^Xg_T^Y|$ & $V_{11}$ & $|f_{11}^{XY}|=2m_X|G|$ & $|g_A^Xg_V^Y|=4m_Y|G|$ \\
spin 1 & $|g_A^Xg_{\widetilde T}^Y|$ & $V_{6+7}$ & $|f_{6+7}^{XY}|=4m_X|G|$ in the DM normalization of Eq.~(5) of Ref.~\cite{Cong2025} & no standard Fig.~9--15 product; this is the extended $g_Ag_{\widetilde T}$ interaction discussed in Sec.~IV.E of Ref.~\cite{Cong2025} \\
spin 1 & $|g_T^Xg_T^Y|$ & $V_3$ in $V_2+V_3$ & $|f_3^{XY}|=4m_X^2|G|$ & $|g_V^Xg_V^Y|_3=|g_p^Xg_p^Y|_3=16m_Xm_Y|G|$ \\
spin 1 & $|g_{\widetilde T}^Xg_{\widetilde T}^Y|$ & $V_3$ & $|f_3^{XY}|=4m_X^2|G|$ & $|g_p^Xg_p^Y|=16m_Xm_Y|G|$ \\
\midrule
spin 2 & $|g_2^Xg_2^Y|$ & $V_1$ & $|f_1^{XY}|={2\over3}m_Xm_Y|G|$ & $|g_V^Xg_V^Y|_{1}={2\over3}m_Xm_Y|G|$ \\
spin 2 & $|g_2^Xg_2^Y|$ & $V_3$ component & $|f_3^{XY}|={m_X^2\over8}|G|$ & $|g_V^Xg_V^Y|_{3}={1\over2}m_Xm_Y|G|$ \\
spin 2 & $|g_2^Xg_2^Y|$ & $V_{4+5}$ & $|f_{4+5}^{XY}|=2m_X^2a_X|G|$ & $\displaystyle |g_V^Xg_V^Y|_{4+5}={2m_X^2a_X\over1/2+m_X/m_Y}|G|$ \\
spin 2 & $|g_2^X\widetilde g_2^Y|$ & $V_{11}$ & $|f_{11}^{XY}|={m_Xm_Y\over4}|G|$ &
$\begin{aligned}[t]
  |g_V^Xg_A^Y|_{11}&={m_Xm_Y\over2}|G|,\\[1pt]
  |g_A^Xg_V^Y|_{11}&={m_Y^2\over2}|G|
\end{aligned}$\,
\footnotemark[1]\\
spin 2 & $|g_2^X\widetilde g_2^Y|$ & exchanged $V_{12,13}$ & $|f_{12+13}^{YX}|=2m_Y(m_X+m_Y)|G|$ & $|g_A^Yg_V^X|_{12+13}=m_Xm_Y|G|$ \\
spin 2 & $|\widetilde g_2^X\widetilde g_2^Y|$ & $V_2$ & $|f_2^{XY}|={m_Xm_Y\over2}|G|$ & $|g_A^Xg_A^Y|={m_Xm_Y\over2}|G|$ \\
spin 2 & $|\widetilde g_2^X\widetilde g_2^Y|$ & $V_3$ & $\displaystyle |f_3^{XY}|={m_X^3m_Y\over2M^2}|G|$ & same full leading $V_2+V_3$ as $|g_A^Xg_A^Y|={m_Xm_Y\over2}|G|$ \\
\bottomrule
\end{tabular}
\footnotetext[1]{Both vertex assignments of the axial coupling in
$f_{11}^{XY}={1\over2}g_V^Xg_A^Y+{m_X\over2m_Y}g_A^Xg_V^Y$ are listed;
Eqs.~\eqref{eq:boundAT} and~\eqref{eq:S2V11bound} use the second.}
\end{table*}

Two features of Table~\ref{tab:newRMPmatch} are worth stressing.  The
spin-1 $g_Tg_V$ and spin-2 $g_2g_2$ interactions each give more than one
standard $V_i$ component, and the effective $g_Vg_V$ rescaling is different
for the different components.  Likewise, the spin-2 $g_2\widetilde g_2$
interaction has a different relative coefficient of $V_{11}$ and
$V_{12,13}$ from a standard vector--axial-vector interaction.  Therefore the
strongest rigorous bound on a complete model requires folding the complete
potential through the response function of the experiment; the relations in
Table~\ref{tab:newRMPmatch} are exact translations of the stated components.

\paragraph{Why exclusion plots are not repeated here.}
Section V of Ref.~\cite{Cong2025} already collects the up-to-date
mass-dependent exclusion regions for the standard physical products:
$g_Ag_V$ in Fig.~9, $g_Ag_A$ in Figs.~10 and 11, $g_Vg_V$ in Fig.~12,
$g_pg_s$ in Fig.~13, $g_pg_p$ in Fig.~14, and $g_sg_s$ in Fig.~15.
Appendix A and Table VI of that review provide the corresponding conversion
to the DM coefficients $f_i^{XY}$.  Thus the combination of those figures
with Table~\ref{tab:newRMPmatch} gives the continuous exclusion curves for
the new products considered here, and reproducing essentially the same
plots would add little information.  Strictly speaking, Ref.~\cite{Cong2025}
does not draw a separate plot for every individual $f_i$; rather, its Table
VI supplies the conversion between $f_i$ and the physical coupling products.
The new TOPE product $g_Ag_{\widetilde T}$ is also an exception: it is
discussed in Sec.~IV.E of Ref.~\cite{Cong2025}, where the relevant
$V_{6,7}$ searches are listed, but it is not one of the standard physical
products plotted in Figs.~9--15.  This is why we retain a small table of
representative $V_{6+7}$ limits below.

For convenience, Table~\ref{tab:RMPplotguide} gives a direct guide from the
figures of Ref.~\cite{Cong2025} to the coupling products in this work.  This
also makes explicit which published exclusion plot should be used when a
continuous mass-dependent limit, rather than one of the representative
numbers below, is required.

\begin{table*}[t]
\caption{Guide to the exclusion plots in Ref.~\cite{Cong2025}.  The third
column lists the principal RMP/Fadeev potential components entering those
plots; the last column gives examples of the new products in this paper that
can be constrained component by component using Table~\ref{tab:newRMPmatch}.
The TOPE $V_{6,7}$ interaction is discussed separately in Sec.~IV.E of
Ref.~\cite{Cong2025}.}
\label{tab:RMPplotguide}
\centering\scriptsize
\renewcommand{\arraystretch}{1.18}
\begin{tabular}{@{}p{0.10\textwidth}p{0.16\textwidth}p{0.29\textwidth}p{0.36\textwidth}@{}}
\toprule
RMP location & Plotted physical product & Principal potential terms & New products accessible through the same component \\
\midrule
Fig.~9 & $g_A^Xg_V^Y$ & $V_{12+13}$, $V_{11}$, $V_{11p+16}$ & $g_A^Xg_T^Y$ through $V_{11}$; $g_2^X\widetilde g_2^Y$ through $V_{11}$ or $V_{12,13}$ \\
Figs.~10,11 & $g_A^Xg_A^Y$ & $V_2$, $V_3$, $V_{4+5}$, $V_8$ & $\widetilde g_2^X\widetilde g_2^Y$ through the full leading $V_2+V_3$ structure; selected $g_2^Xg_2^Y$ components \\
Fig.~12 & $g_V^Xg_V^Y$ & $V_1$, $V_2+V_3$, $V_{4+5}$ & $g_T^Xg_V^Y$, $g_T^Xg_T^Y$, and $g_2^Xg_2^Y$ through the indicated components \\
Fig.~13 & $g_p^Xg_s^Y$ & $V_{9+10}$, $V_{14q+15}$ & $g_{\widetilde T}^Xg_V^Y$ through $V_{9+10}$; the complete higher-order response is required for $V_{14q+15}$ \\
Fig.~14 & $g_p^Xg_p^Y$ & $V_3$ & $g_T^Xg_T^Y$, $g_{\widetilde T}^Xg_{\widetilde T}^Y$, and the $V_3$ components of the corresponding spin-2 interactions \\
Fig.~15 & $g_s^Xg_s^Y$ & $V_1$ and relativistic $V_{4+5}$ corrections & the $V_1$ component of $g_2^Xg_2^Y$ (and, component-wise, $V_{4+5}$) \\
Table IV & $\operatorname{Re}C_X\operatorname{Re}C_Y/\Lambda^4$ and $\operatorname{Im}C_X\operatorname{Im}C_Y/\Lambda^4$ for a strictly massless spin-1 boson & $V_3$ & strict-long-range limits on $g_T^Xg_T^Y$ and $g_{\widetilde T}^Xg_{\widetilde T}^Y$; operator-level input for the small-$M$ spin-2 translations in Tables~\ref{tab:TT_TtildeTtilde_asymptotic},\ref{tab:g2g2_asymptotic} and  \ref{tab:g2tildeg2_asymptotic} \\
\bottomrule
\end{tabular}
\end{table*}

\subsection{Light- and heavy-mediator combinations}
\label{subsec:asymptoticbounds}

Let $Q$ denote the characteristic momentum transfer of a given experiment.
For $M\ll Q$ the Yukawa propagator approaches the long-range form, whereas
for $M\gg Q$
\begin{equation}
 {1\over \bq^2+M^2}={1\over M^2}-{\bq^2\over M^4}+\cdots .
\end{equation}
Consequently a short-range experiment constrains a Wilson coefficient built
from the coupling product divided by $M^2$, with additional derivatives when
the vertices themselves contain momenta. These are
scaling statements; a numerical contact bound is experiment dependent
because the matrix element of the delta function or its derivatives depends on the source geometry or atomic/molecular wave function. Appendix~A tabulates the resulting limits for each coupling combination.

The spin-2 axial-tensor--axial-tensor channel (Table~\ref{tab:g2tildeg2_asymptotic}) has a direct analogue in the massive spin-1 axial--axial
potential. For fixed couplings, the coefficient of the longitudinal
$V_3|_{AA}$ contribution scales as $1/M^2$ and therefore becomes
strongly enhanced as $M\to0$. In the small-$M$ regime, experimental
limits on this contribution are naturally expressed in terms of the
finite combination $g_A^Xg_A^Y\lambda^2$, equivalently
$g_A^Xg_A^Y/M^2$, as discussed in Ref.~\cite{Cong2025}. Through the exact leading mapping
$g_A^Xg_A^Y=-(m_Xm_Y/2)\widetilde g_2^X\widetilde g_2^Y$, the corresponding
spin-2 quantity is
\begin{equation}
 \left|{\widetilde g_2^X\widetilde g_2^Y\over M^2}\right|
 ={2\over m_Xm_Y}
 \left|{g_A^Xg_A^Y\over M^2}\right|.
\label{eq:smallMspin2AA}
\end{equation}
This statement concerns a small but nonzero massive mediator.  It must not be
replaced by the bounds for a genuinely massless axial vector, for which the
longitudinal $1/M^2$ term is absent.

For all component maps whose matching factor $K_{XY}$ contains only fermion
masses, a complete RMP exclusion curve $G_{\rm RMP}^{\rm lim}(M)$ immediately
gives
\begin{align}
 |G_{\rm new}|_{\rm lim}&=K_{XY}|G_{\rm RMP}|_{\rm lim},\nonumber\\
 \left|\frac{G_{\rm new}}{M^2}\right|_{\rm lim}
 &=K_{XY}\left|\frac{G_{\rm RMP}}{M^2}\right|_{\rm lim}.
\label{eq:curveconversion}
\end{align}
Thus the same translation table applies to both the long-range plateau and
the contact-normalized large-$M$ limit.  Equation~\eqref{eq:curveconversion},
together with Figs.~9--15 of Ref.~\cite{Cong2025}, is more informative than a
second set of exclusion plots in this paper.

\subsection{Spin-1 coupling products}

The mapping above permits existing searches for individual $V_i$ structures
to be translated into limits on the new spin-1 coupling products.  The tables in Appendix~A quote limits on these products. Sec.~\ref{Appendix1} provides limits in the small and large mediator mass $M$. These are quoted explicitly from the source papers. Sec.~\ref{Appendix2} provides representative limits obtained by digitizing published exclusion curves using WebPlotDigitizer and translating a single representative point near the strongest sensitivity region. Therefore, the values presented are approximate. A complete and accurate
determination of the constraints would require digitising and translating the entire and more up-to-date exclusion curves, which should be taken from
Ref.~\cite{Cong2025} as described in Sec.~\ref{subsec:RMPdictionary}. Limits in both of these sections
are component-wise \emph{single-operator translations}, in which the coupling combination under consideration is assumed to be the only non-zero interaction. However, it is important to note coupling product can
generate several $V_i$ structures simultaneously.

For the RMP/DM normalization of Table~\ref{tab:RMPpotentials}, and with
fermion $X$ carrying the displayed spin in $V_{4+5}$, the relevant component
relations are
\begin{align}
|g_T^Xg_V^Y|&=\frac{m_r}{4m_X^2}|f_{4+5}^{XY}|
\xrightarrow[m_Y\gg m_X]{}\frac{|f_{4+5}^{XY}|}{4m_X},
\label{eq:boundVT}\\
|g_{\widetilde T}^Xg_V^Y|&=\frac{|g_p^Xg_s^Y|}{4m_X},
\label{eq:boundVPT}\\
|g_A^Xg_T^Y|&=\frac{|g_A^Xg_V^Y|}{4m_Y},
\label{eq:boundAT}\\
|g_A^Xg_{\widetilde T}^Y|&=\frac{|f_{6+7}^{XY}|_{\rm RMP}}{4m_X},
\label{eq:boundAPTRMP}\\
|g_T^Xg_T^Y|=|g_{\widetilde T}^Xg_{\widetilde T}^Y|
&=\frac{|g_p^Xg_p^Y|}{16m_Xm_Y}.
\label{eq:boundTTpp}
\end{align}
The TOPE numbers quoted below follow the legacy experimental
$f_{6+7}^{\rm exp}$ convention.  In the notation of that table,
$|g_Ag_{\widetilde T}|=|f_{6+7}^{\rm exp}|/(2m_r)$, so the conversion to the
RMP/DM normalization used above is
\begin{equation}
 |f_{6+7}^{XY}|_{\rm RMP}={2m_X\over m_r}|f_{6+7}|_{\rm exp}.
\label{eq:f67conventions}
\end{equation}
Here $m_r=m_1m_2/(m_1+m_2)$ is the reduced mass.  All numerical
conversions below use $m_e=0.510\,998\,950$~MeV, $m_n=939.565\,421$~MeV,
$m_p=938.272\,088$~MeV, and $m_\mu=105.658\,375$~MeV.  The $e$--$n$ entry Ji
\emph{et al.} in
the TOPE table are sensitivity projections from Ref.~\cite{Ji2017}, which is
a proposal rather than a measurement; they are labelled accordingly.

The same $V_3$ operator limits can be used to display particularly simple
small-mass combinations for the massive spin-2 theory.  Using
Eqs.~\eqref{eq:mapS2TT} and \eqref{eq:mapS2PTPT}, Tables~\ref{tab:g2g2_asymptotic} and \ref{tab:g2tildeg2_asymptotic}
show (i) a component-wise limit on $g_2^Xg_2^Y$ obtained from the common
$V_3$ term and (ii) the finite longitudinal combination
$\widetilde g_2^X\widetilde g_2^Y/M^2$.  The latter is the natural quantity
when $M$ is small but nonzero.  These translations use the experimental
operator bounds summarized by the RMP; they do \emph{not} identify a
strictly massless paraphoton with a massless spin-2 field.

Reference~\cite{Cong2025} also emphasizes that no experimental or
observational constraint on the complete massless tensor--pseudotensor
$V_{14q+15}$ interaction was available at the time of that review.  Hence a
model-independent strict-$M=0$ bound on
$g_{\widetilde T}^Xg_T^Y$ cannot be obtained by using a $V_{15}$ limit alone.

We therefore quote no numerical limit on the mixed
$g_{\widetilde T}g_T$ product: the required complete
$q^2V_{14}\pm V_{15}$ response has not been constrained independently.

\subsection{Spin-2 reinterpretations}
\label{subsec:S2bounds}

The minimal spin-2 interactions of Eq.~\eqref{eq:Lspin2} are $T$ even.
Consequently the $P,T$-odd constraints in Tables~\ref{tab:S1VPT} and
\ref{tab:S1VPTEDM}, and the TOPE constraints in Table~\ref{tab:S1APT}, have
no counterpart in this minimal spin-2 operator basis.  The remaining shared
$V_i$ components can be reinterpreted directly, provided one keeps clear that
the complete spin-2 interaction generally contains additional structures.

First consider $V_{4+5}$.  For an unpolarized fermion $b$ and a polarized
fermion $a$, define
\begin{equation}
 A_a=\frac12+\frac{m_b}{3m_a}.
\end{equation}
Using $\bP=m_r\bv$ and the classical source--sensor limit of the
anticommutator in Eq.~\eqref{eq:spin2SOr}, comparison with the standard
$V_{4+5}$ normalization gives
\begin{equation}
 |g_{2a}g_{2b}|=\frac{|f_{4+5}^{ab}|}{2m_a^2 A_a}.
\label{eq:S2V45bound}
\end{equation}
Equation~\eqref{eq:S2V45bound} uses the exact RMP Appendix-A normalization.
The form based on the commonly used macroscopic $f_\perp$
normalization differs by the factor $m_r/m_a$, which is negligible for $e$--$N$. 

The $V_2+V_3$ component of the spin-2 $g_2g_2$ potential and the spin-1
$g_Tg_T$ potential are related by Eq.~\eqref{eq:mapS2TT}.  Likewise the
$V_3$ component of the spin-2 $\widetilde g_2\widetilde g_2$ potential and
the spin-1 $g_{\widetilde T}g_{\widetilde T}$ potential are related by
Eq.~\eqref{eq:mapS2PTPT}.  Thus
\begin{align}
 |g_{21}g_{22}|&=32|g_T^1g_T^2|,
\label{eq:S2TTbound}\\
 |\widetilde g_{21}\widetilde g_{22}|_{V_3}
 &=\frac{8M^2}{m_1m_2}|g_{\widetilde T}^1g_{\widetilde T}^2|,
\qquad M=\frac{\hbar c}{\lambda}.
\label{eq:S2PTPTbound}
\end{align}
The second relation is explicitly a $V_3$-component translation; the full
spin-2 axial-tensor--axial-tensor potential also contains $V_2$.

Finally, the common $V_{11}$ component of the spin-1 axial--tensor and
spin-2 tensor--axial-tensor interactions gives
\begin{align}
 |g_{21}\widetilde g_{22}|
 &=\frac{8}{m_2}\,|g_A^1g_T^2|
\nonumber\\
 &=\frac{2}{m_2^2}\,|g_A^1g_V^2|.
\label{eq:S2V11bound}
\end{align}
The particle ordering is the one in Eq.~\eqref{eq:mapS2V11}: a source limit
on $g_A^1g_V^2$ maps to ordinary spin-2 coupling on particle 1 and
axial-tensor spin-2 coupling on particle 2.  The complete spin-2 mixed
potential also contains an unsuppressed $V_{12,13}$ contribution.

For completeness, Eq.~\eqref{eq:mapS2AAexact} gives an exact leading-order
functional rescaling between the spin-2
$\widetilde g_2\widetilde g_2$ interaction and ordinary spin-1 axial--axial
exchange whenever the experimental analysis uses the full $V_2+V_3$ Proca
potential.  This provides an independent route to spin-2 bounds beyond the
component translations tabulated above.  If a publication constrains only
$V_2$ or only $V_3$, the full-model limit should instead be obtained by
recomputing the experimental response to the complete spin-2 potential.

\section{Conclusion}
\label{sec:conclusion}

We have derived finite-range two-fermion potentials for a massive spin-1 mediator carrying simultaneously vector/axial-vector and tensor/pseudotensor interactions.  The mixed vertices generate operator combinations not present when the vector and dipole sectors are treated separately.  In particular,
\begin{equation}
 g_Ag_{\widetilde T}\quad\Longrightarrow\quad V_{6,7},
 \qquad P=+1,\quad T=-1,
\end{equation}
providing a tree-level single-boson mechanism for the TOPE spin--velocity interaction in the standard sixteen-potential basis.  The results given here include both vertex orderings and the distributional contact terms required for short-range and atomic-scale applications.

We also extended the analysis to a massive spin-2 field.  For the conventional conserved energy--momentum coupling the longitudinal propagator terms vanish exactly.  A symmetric axial-tensor fermion current is not conserved for massive fermions, so its exchange probes the longitudinal spin-2 helicities.  The mixed tensor--axial-tensor interaction is $P$ odd but $T$ even, while the tensor--tensor and axial-tensor--axial-tensor interactions are $P,T$ even.  Thus no $T$-odd force is produced by the minimal spin-2 operator basis considered here.  The heavy-mediator limit yields local derivative operators, whereas a genuine massless graviton is restricted by gauge invariance to conserved symmetric sources and gives the usual $P,T$-even gravitational interaction.

We have also mapped the new interactions onto the sixteen-potential Dobrescu--Mocioiu basis and used this mapping to obtain limits on the new coupling products, both in the small- and large- mediator-mass $M$ regime, where the constraints are quoted directly from the source papers, and at a representative finite masses obtained by digitising the published exclusion curves.  Among the new tensor/pseudotensor and spin-2 channels derived explicitly here, $V_{16}$ is not generated at the order retained. The conventional axial-vector--vector Proca interaction can generate $V_{16}$ at higher order~\cite{Cong2025}.  Where spin-1 and spin-2 exchange share an identical operator and radial function, we supplied explicit coefficient maps and corresponding spin-2 component bounds.  These results provide a unified 
starting point for interpreting macroscopic fifth-force measurements, precision spin experiments, and atomic or molecular tests in terms of massive spin-1 and spin-2 mediators.  Dedicated analyses incorporating all operators generated by a given relativistic coupling combination will be needed for rigorous full-model bounds.

\vspace{3mm}
\textit{Acknowledgments.} -- This work was supported by the Australian Research Council Grant No.\ DP230101058.

\appendix
\section {Constraints on coupling combinations}
This appendix presents the constraints on the spin-1 and spin-2 interaction constants considered in this work. The bounds are obtained by matching components of the potentials derived above to existing experimental constraints, assuming in each case that the coupling combination under consideration is the only nonzero new interaction. We first present the limits for small- and large-mediator-mass regimes, where available, followed by representative constraints at finite mediator masses. The representative limits are obtained by digitising the published exclusion curves and are intended to illustrate the sensitivity to the newly derived coupling combinations; they should therefore be regarded as approximate rather than as a complete reanalysis of the experimental data.

\subsection{Constraints in the regimes of large and small masses M of the mediator.}
\label{Appendix1}

 In the case of a small mass M of the mediator, the  Yukawa exponent $\exp(-Mr) \approx 1 $,  and the constraints on the product of the interaction constants $g_ig_k$  usually do not depend on $M$. In experiments with atoms and small molecules, $r \sim a_B$, so we consider the mass regime $M a_B \ll 1$, giving $M \ll$ 4 keV,  where $a_B$ is the Bohr radius. In the opposite, large-mass regime, we assume the contact limit, $Y_M(r)\rightarrow\delta^3(\br)/M^2$, so constraints are usually expressed in terms of $g_i g_k/M^2$. In atomic experiments this typically corresponds to $M a_B/Z \gg 1$, giving $M \gg$ 4 $Z$ keV, where $Z$ is the nuclear charge.  These are characteristic atomic estimates, since the relevant momentum scale is experiment- and observable - dependent.

Limits for small and large mediator masses on the spin-1
vector--tensor coupling $|g_V^1 g_T^2|$ are presented in the Table \ref{tab:VT_asymptotic}. The $M\to0$ limits
are obtained by matching the common $V_3$ component of the
interactions to the tensor--tensor/pseudotensor--pseudotensor
constraints for a massless paraphoton reported in Table~IV of
Ref.~\cite{Cong2025}, using
$|g_V^1g_T^2|
=4v_h^2m_1
|\mathrm{Re}(C_1)\mathrm{Re}(C_2)/\Lambda^4|$,
with normalisation $g_T=\frac{v_h}{\Lambda^2}\mathrm{Re}(C)$, and $v_h=246~\mathrm{GeV}$.
The large-$M$ (contact) limit is obtained from the
pseudoscalar--pseudoscalar $V_3$ constraint using
$|g_V^1g_T^2|/M^2
=|g_p^1g_p^2|/(4m_2M^2)$. This is a component-wise $V_3$ reinterpretation. The additional $V_2$ contribution present in the full combination $(V_2+V_3)|_{VT}$ is not included and may become important in the short-range (contact) regime. The complete vector--tensor interaction also contains the momentum-suppressed $\bq^2V_1$ and $V_{4,5}$ terms, which are likewise omitted here but may contribute to the response of a particular experiment. 

Limits for small and large mediator masses on the spin-1
vector--pseudotensor coupling $|g_{\widetilde T}^{1}g_V^{2}|$,
obtained from constraints on the $V_{9+10}$ interaction using
$|g_{\widetilde T}^{1}g_V^{2}|
=|g_p^{1}g_s^{2}|/(4m_1)$, are presented in the Table \ref{tab:gTtildeV_limits}.
The small-$M$ limits apply in the long-range regime,
$M\ll 1~\mathrm{keV}$, while the large-$M$ limits apply in the
contact regime, $M\gg 1~\mathrm{GeV}$; see Table~III of
Ref.~\cite{Prosnyak2023}. These limits are component-wise $V_{9+10}$ reinterpretations. The complete vector--pseudotensor interaction also contains momentum-suppressed $\bq^2V_{14}\pm V_{15}$ terms, which are not included in this translation.

Limits for small and large mediator masses on the spin-1
axial--tensor coupling $|g_A^1g_T^2|$ are presented in the Table \ref{tab:gAgT_limits}. The small-$M$ limit is
obtained from the $V_{11}$ constraint using
$|g_A^1g_T^2|=|g_A^1g_V^2|/(4m_2)$ and applies for
$M\lesssim 0.1\,\mu\mathrm{eV}$.
This is a component-wise $V_{11}$ reinterpretation.
The complete axial--tensor interaction also contains the
momentum-suppressed $\bq^2V_{12+13}$ contribution, which is not
included in this translation. The quoted $e$--$e$ constraint
is a $1\sigma$ limit from Ref.~\cite{Heckel2013}.
No corresponding large-$M$ $V_{11}$ limit was identified.

Constraints on the spin-1 axial--pseudotensor coupling
$g_A g_{\widetilde T}$, obtained by translating the dimensionless
Conti--Khriplovich constants $\beta$~\cite{Conti1992}
via the dictionary derived here,
$|g_A g_{\widetilde T}|=(\pi/m_p)|\beta|
=3.35|\beta|~{\rm GeV}^{-1}$ are presented in the Table \ref{tab:gAgTtilde_limits}.
Here $m_p$ is not a dynamical fermion mass but the reference mass
introduced in Ref.~\cite{Conti1992} through the factor $1/(2m)$,
with $m$ conventionally fixed to the proton mass so that $\beta$
is dimensionless.
The quoted $\beta$ bounds are indirect constraints inferred from
electron and neutron EDM limits through two-loop $Z$-exchange
diagrams. Their mediator-mass dependence enters through
$\ln(\Lambda^2/M_>^2)$, with $M_>=\max(M,M_Z)$.
The underlying leading-log expression is independent of $M$ below
$M_Z$ and depends only logarithmically on $M$ above $M_Z$; for
their numerical estimates Ref.~\cite{Conti1992} conservatively
set this logarithm to unity.
For the nonidentical-fermion channels, the fourth column shows the
same translated coupling constraint divided by $M_Z^2$, evaluated
at $M=M_Z=91.19~{\rm GeV}$. It is included only as a reference
contact-form normalization and is not an independent high-mass
asymptote of a Yukawa-force exclusion curve. Equivalently, Ref.~\cite{Conti1992} parametrizes the local
four-fermion interaction by
$4\pi\beta/M^2=(G_F/\sqrt{2})\,\widetilde q$, which gives
$|g_A g_{\widetilde T}|/M^2
=G_F|\widetilde q|/(4\sqrt{2}\,m_p)
=2.20\times10^{-6}|\widetilde q|~{\rm GeV}^{-3}$.
The $\widetilde q$ limits quoted in Ref.~\cite{Conti1992} were
evaluated for $M\sim100m_p$ and rounded at the order-of-magnitude
level, and are therefore used only as a cross-check rather than as
the numerical input to the fourth column.
For identical fermions the local $e$--$e$ TOPE interaction
vanishes because the scattering contribution is cancelled by
the exchange contribution~\cite{Conti1992};
therefore, no corresponding contact entry is given.
The nucleon-level entries follow from quark-level estimates of
the nucleon matrix elements and therefore carry an additional
order-of-magnitude uncertainty. The first index of $\beta$ denotes
the pseudotensor (derivative) vertex. The EDM-derived bounds quoted here are conditional on assumptions about the short-distance realization of the TOPE interaction. A subsequent effective field theory analysis showed that the connection between EDM and direct TOPE constraints depends on the ultraviolet realization of parity violation \cite{Kurylov2001}. We therefore treat the Conti--Khriplovich bounds as conditional constraints rather than fully model-independent constraints.

Limits for small and large mediator masses on the spin-1
tensor--tensor and pseudotensor--pseudotensor couplings,
$|g_T^1g_T^2|$ and
$|g_{\widetilde T}^1g_{\widetilde T}^2|$, are presented in the Table \ref{tab:TT_TtildeTtilde_asymptotic}.
The small-$M$ limits are obtained from the massless-paraphoton
constraints in Table~IV of Ref.~\cite{Cong2025}, using
$|g_T^1g_T^2|
=v_h^2|\mathrm{Re}(C_1)\mathrm{Re}(C_2)/\Lambda^4|$
and
$|g_{\widetilde T}^1g_{\widetilde T}^2|
=v_h^2|\mathrm{Im}(C_1)\mathrm{Im}(C_2)/\Lambda^4|$,
with normalisation $g_T=\frac{v_h}{\Lambda^2}\mathrm{Re}(C)$, $g_{\widetilde T}= \frac{v_h}{\Lambda^2}\mathrm{Im}(C)$, and $v_h=246~\mathrm{GeV}$.
The large-$M$ limits are obtained from the
pseudoscalar--pseudoscalar $V_3$ constraint using
$|g_T^1g_T^2|/M^2
=|g_{\widetilde T}^1g_{\widetilde T}^2|/M^2
=|g_p^1g_p^2|/(16m_1m_2M^2)$.
For $g_Tg_T$, the complete tensor--tensor potential also contains a $V_2$ contribution, which is not included in this translation
and may contribute to the response of a particular experiment. For
$g_{\widetilde T}g_{\widetilde T}$, the interaction is a pure
$V_3$ structure at the order retained.
The resulting $V_3$-based numerical limits are identical for the
two couplings. 

Limits for small and large mediator masses on the
tensor--tensor coupling $|g_2^1g_2^2|$ of a massive spin-2
($S=2$) mediator are presented in the Table \ref{tab:g2g2_asymptotic}. The limits are obtained from the common
$V_3$ component of the spin-2 tensor--tensor and spin-1
tensor--tensor interactions, using
$|g_2^1g_2^2|=32|g_T^1g_T^2|$.
The small-$M$ limits are obtained by rescaling the corresponding
tensor--tensor/pseudotensor--pseudotensor constraints for a
massless paraphoton reported in Table~IV of
Ref.~\cite{Cong2025}. This is a translation based only on the
common $V_3$ interaction structure and should not be interpreted
as the strictly massless limit of the spin-2 theory; the mediator
considered here remains a massive Fierz--Pauli spin-2 field.
The large-$M$ (contact) limit is obtained using
$|g_2^1g_2^2|/M^2
=32|g_T^1g_T^2|/M^2$.
This is a component-wise reinterpretation: the complete spin-2 tensor–tensor potential also contains the central $V_1$ term, the $V_2$ component accompanying $V_3$, and $V_{4,5}$ terms, which are not included in this $V_3$-only translation.

Limits for small and large mediator masses on the mixed
tensor--axial-tensor coupling $|\widetilde g_2^1g_2^2|$ of a
massive spin-2 ($S=2$) mediator are presented in the Table \ref{tab:g2g2tilde_asymptotic}. The limits are obtained by
matching the common $V_{12+13}$ component of the spin-2
$g_2\widetilde g_2$ interaction to the axial--vector interaction
of a massive spin-1 mediator, using
$|\widetilde g_2^1g_2^2|
=|g_A^1g_V^2|/(m_1m_2)$.
The $^{133}$Cs, $^{174}$Yb, and $^{205}$Tl rows are obtained from
atomic parity-nonconservation constraints on
$g_A^e g_V^N$, where
$g_V^N=(Ng_V^n+Zg_V^p)/A$, and therefore correspond to effective
$e$--$N$ couplings.
The $^{133}$Cs anapole row instead constrains
$g_A^p g_V^N$ and therefore corresponds to a $p$--$N$ coupling.
For the numerical conversion we use
$m_N\simeq0.939\,\mathrm{GeV}$.
The small-$M$ column is obtained from the long-range limits quoted
in Ref.~\cite{Dzuba2017}, whereas the large-$M$ column gives the
contact-normalized limits.
These are component-wise reinterpretations based only on the
shared $V_{12+13}$ interaction structure and should not be
interpreted as the strictly massless limit of the spin-2 theory;
the mediator considered here remains a massive Fierz--Pauli
spin-2 field. The complete spin-2 $g_2\widetilde g_2$
interaction also contains a $V_{11}$ contribution, which is not
included in this translation.

Constraints on the axial-tensor--axial-tensor coupling
$\widetilde g_2^1\widetilde g_2^2$ of a massive spin-2 ($S=2$)
mediator for small and large mediator masses are presented in the Table \ref{tab:g2tildeg2_asymptotic}. Both regimes constrain the combination
$|\widetilde g_2^1\widetilde g_2^2|/M^2$, but for different physical reasons. In the small-$M$ regime, the $1/M^2$ dependence arises from the longitudinally enhanced $V_3$ component of the massive spin-2 interaction; the quoted small-$M$ limits $|\widetilde g_2^1\widetilde g_2^2|/M^2$ are therefore component-wise $V_3$ reinterpretations. In the large-$M$ regime, the $V_3$ contribution is $\mathcal{O}(M^{-4})$, while the leading $V_2$ term acquires the usual $1/M^2$ dependence in the contact limit; the quoted large-$M$ limits $|\widetilde g_2^1\widetilde g_2^2|/M^2$ are therefore component-wise $V_2$ reinterpretations. 
The positronium, muonium and helium entries (rows 2, 3 and 4) are obtained from axial--axial
constraints using the mapping
$|\widetilde g_2^1\widetilde g_2^2|/M^2
=2|g_A^1g_A^2|/(M^2(m_1m_2))$.
Other small-$M$ entries are component-wise $V_3$ reinterpretations of pseudotensor--pseudotensor constraints using
$|\widetilde g_2^1\widetilde g_2^2|/M^2
=8|g_{\widetilde T}^1g_{\widetilde T}^2|/(m_1m_2)$. Only the corresponding $V_3$ or $V_2$ component is included in
each translation; other terms in the complete spin-2
axial-tensor--axial-tensor interaction may contribute to the response of a particular experiment.
The small-$M$ results apply to a massive Fierz--Pauli mediator whose range is large compared with the characteristic length scale of the experiment; they should not be interpreted as constraints on a strictly massless spin-2 mediator. For $N$, the neutron mass is used in the numerical conversion. 

\begin{table}[h]
\centering
\caption{Limits for small and large mediator masses on the spin-1
vector--tensor coupling $|g_V^1 g_T^2|$. An em dash indicates that no corresponding small- or
large-$M$ limit is quoted.}
\label{tab:VT_asymptotic}
\begin{ruledtabular}
\begin{tabular}{cccc}
Fermion pair &
\begin{tabular}{c}
Small-$M$ limit\\
$|g_V^1g_T^2|$($\mathrm{GeV}^{-1}$)
\end{tabular}
&
\begin{tabular}{c}
Large-$M$ limit\\
$|g_V^1g_T^2|/M^2$($\mathrm{GeV}^{-3}$)
\end{tabular}
&
Ref. \\
\hline
$e-e$     & $2.72\times10^{-14}$ & ---                 & \cite{Terrano2015} \\
$e-e^+$   & $3.83\times10^{-3}$ & ---                 & \cite{Fadeev2022} \\
$e-\mu^+$ & $4.95\times10^{-8}$ & ---                 & \cite{Fadeev2022} \\
$e-\bar{p}$     & $4.95\times10^{-4}$ & --- & \cite{Ficek2018} \\
$e-n$     & $4.58\times10^{-15}$ & ---                 & \cite{Almasi2020} \\
$p-p$     & $7.72\times10^{-5}$ & ---                 & \cite{Ramsey1979} \\
$n-n$     & $1.93\times10^{-9}$ & ---                 & \cite{Vasilakis2009} \\
$e-p$     & ---  & $4.80\times10^{2}$                & \cite{Fadeev2022}\\
\end{tabular}
\end{ruledtabular}
\end{table}

\begin{table}[!tb]
\centering
\caption{Limits for small and large mediator masses on the spin-1
vector--pseudotensor coupling $|g_{\widetilde T}^{1}g_V^{2}|$.}

\label{tab:gTtildeV_limits}
\begin{tabular}{c c c c}
\hline\hline
Fermion pair &
Small-$M$ limit &
Large-$M$ limit &
Ref. \\
&
$|g_{\widetilde T}^1g_V^2|$ [GeV$^{-1}$] &
$|g_{\widetilde T}^1g_V^2|/M^2$ [GeV$^{-3}$] &
\\
\hline
$e-e$
& $1.08\times10^{-17}$
& $8.32\times10^{-9}$
& \cite{Prosnyak2023} \\

$e-N$
& $5.38\times10^{-18}$
& $1.57\times10^{-12}$
& \cite{Prosnyak2023} \\
\hline\hline
\end{tabular}
\end{table}

\begin{table}[!tb]
\centering
\caption{Limits for small and large mediator masses on the spin-1
axial--tensor coupling $|g_A^1g_T^2|$. An em dash indicates that no corresponding small- or
large-$M$ limit is quoted.}
\label{tab:gAgT_limits}
\begin{tabular}{c c c c}
\hline\hline
Fermion pair &
Small-$M$ limit &
Large-$M$ limit &
Ref. \\
&
$|g_A^1g_T^2|$ [GeV$^{-1}$] &
$|g_A^1g_T^2|/M^2$ [GeV$^{-3}$] &
\\
\hline
$e$--$e$
& $5.87\times10^{-26}$
& --- &
\cite{Heckel2013} \\
\hline\hline
\end{tabular}
\end{table}

\begin{table}[!tb]
\centering
\caption{Constraints on the spin-1 axial--pseudotensor coupling
$g_A g_{\widetilde T}$, obtained by translating the dimensionless
Conti--Khriplovich constants $\beta$~\cite{Conti1992}. An em dash indicates that no corresponding small- or
large-$M$ limit is quoted.}
\label{tab:gAgTtilde_limits}

\begin{tabular}{c c c c c}
\hline\hline
Fermion pair &
$|\beta|$ &
$|g_{\widetilde T}^1g_A^2|$ &
$|g_{\widetilde T}^1g_A^2|/M_Z^2$ &
Ref. \\
&
&
[GeV$^{-1}$] &
[GeV$^{-3}$] &
\\
\hline

$e$--$e$
& $1\times10^{-7}$
& $3.3\times10^{-7}$
& ---
& \cite{Conti1992} \\

$e$--$N$, deriv.\ on $e$
& $3\times10^{-8}$
& $1.0\times10^{-7}$
& $1.2\times10^{-11}$
& \cite{Conti1992} \\

$N$--$e$, deriv.\ on $N$
& $1\times10^{-6}$
& $3.3\times10^{-6}$
& $4.0\times10^{-10}$
& \cite{Conti1992} \\

$N$--$N$
& $3\times10^{-7}$
& $1.0\times10^{-6}$
& $1.2\times10^{-10}$
& \cite{Conti1992} \\

\hline\hline
\end{tabular}
\end{table}

\begin{table}[!tb]
\centering
\caption{Limits for small and large mediator masses on the spin-1
tensor--tensor and pseudotensor--pseudotensor couplings,
$|g_T^1g_T^2|$ and
$|g_{\widetilde T}^1g_{\widetilde T}^2|$, respectively. An em dash indicates that no corresponding
small- or large-$M$ limit is quoted.}

\label{tab:TT_TtildeTtilde_asymptotic}
\begin{ruledtabular}
\begin{tabular}{cccc}
Fermion pair &
\begin{tabular}{c}
Small-$M$ limit\\
$|g_T^1g_T^2|$\\$|g_{\widetilde T}^1g_{\widetilde T}^2|$\\
($\mathrm{GeV}^{-2}$)
\end{tabular}
&
\begin{tabular}{c}
Large-$M$ limit\\
$|g_T^1g_T^2|/M^2$\\
$|g_{\widetilde T}^1g_{\widetilde T}^2|/M^2$\\
($\mathrm{GeV}^{-4}$)
\end{tabular}
&
Ref. \\
\hline
$e-e$     & $1.33\times10^{-11}$ & ---                 & \cite{Terrano2015} \\
$e-e^+$   & $1.88$                & ---                 & \cite{Fadeev2022} \\
$e-\mu^+$ & $2.42\times10^{-5}$  & ---                 & \cite{Fadeev2022} \\
$e-\bar{p}$     & $2.42\times10^{-1}$  & --- & \cite{Ficek2018} \\
$e-n$     & $2.24\times10^{-12}$ & ---                 & \cite{Almasi2020} \\
$p-p$     & $2.06\times10^{-5}$  & ---                 & \cite{Ramsey1979} \\
$n-n$     & $5.14\times10^{-10}$ & ---                 & \cite{Vasilakis2009} \\
$e-p$     & ---  & $2.35\times10^{5}$ & \cite{Fadeev2022} \\
\end{tabular}
\end{ruledtabular}
\end{table}

\begin{table}[!tb]
\centering
\caption{Limits for small and large mediator masses on the
tensor--tensor coupling $|g_2^1g_2^2|$ of a massive spin-2
($S=2$) mediator. An em dash indicates that no corresponding small- or large-$M$
limit is quoted.}
\label{tab:g2g2_asymptotic}
\begin{ruledtabular}
\begin{tabular}{cccc}
Fermion Pair &
\begin{tabular}{c}
Small-$M$ limit\\
$|g_2^1g_2^2|$\\
(GeV$^{-2}$)
\end{tabular}
&
\begin{tabular}{c}
Large-$M$ limit\\
$|g_2^1g_2^2|/M^2$\\
(GeV$^{-4}$)
\end{tabular}
&
Ref. \\
\hline
$e-e$     
& $4.26\times10^{-10}$ 
& --- 
& \cite{Terrano2015} \\

$e-e^+$   
& $6.02\times10^{1}$  
& --- 
& \cite{Fadeev2022} \\

$e-\mu^+$ 
& $7.74\times10^{-4}$ 
& --- 
& \cite{Fadeev2022} \\

$e-\bar{p}$ 
& $7.74$ 
& --- 
& \cite{Ficek2018} \\

$e-n$     
& $7.17\times10^{-11}$ 
& --- 
& \cite{Almasi2020} \\

$p-p$     
& $6.59\times10^{-4}$ 
& --- 
& \cite{Ramsey1979} \\

$n-n$     
& $1.64\times10^{-8}$ 
& --- 
& \cite{Vasilakis2009} \\

$e-p$     
& --- 
& $7.52\times10^{6}$ 
& \cite{Fadeev2022} \\
\end{tabular}
\end{ruledtabular}
\end{table}

\begin{table}[!tb]
\centering
\caption{Limits for small and large mediator masses on the mixed
tensor--axial-tensor coupling $|\widetilde g_2^1g_2^2|$ of a
massive spin-2 ($S=2$) mediator.}
\label{tab:g2g2tilde_asymptotic}
\begin{ruledtabular}
\begin{tabular}{cccc}
Fermion pair &
\begin{tabular}{c}
Small-$M$\\
$|\widetilde g_2^1 g_2^2|$\\
(GeV$^{-2}$)
\end{tabular}
&
\begin{tabular}{c}
Large-$M$\\
$|\widetilde g_2^1 g_2^2|/M^2$\\
(GeV$^{-4}$)
\end{tabular}
&
Ref. \\
\hline

$^{133}$Cs ($\widetilde g_2^eg_2^N$)
& $6.46\times10^{-11}$
& $8.13\times10^{-5}$
& \cite{Dzuba2017} \\

$^{174}$Yb ($\widetilde g_2^eg_2^N$)
& $2.92\times10^{-9}$
& $2.29\times10^{-3}$
& \cite{Dzuba2017} \\

$^{205}$Tl ($\widetilde g_2^eg_2^N$)
& $7.50\times10^{-10}$
& $3.13\times10^{-4}$
& \cite{Dzuba2017} \\

$^{133}$Cs anapole ($\widetilde g_2^pg_2^N$)
& $6.81\times10^{-8}$
& $2.61\times10^{-5}$
& \cite{Dzuba2017} \\

\end{tabular}
\end{ruledtabular}
\end{table}

\begin{table}[!tb]
\centering
\caption{Limits on the axial-tensor--axial-tensor coupling
$\widetilde g_2^1\widetilde g_2^2$ of a massive spin-2 ($S=2$)
mediator for small and large mediator masses. An em dash indicates no corresponding limit is quoted.}
\label{tab:g2tildeg2_asymptotic}

\begin{ruledtabular}
\begin{tabular}{ccccc}
Fermion Pair &
\begin{tabular}{c}
Small-$M$\\
$|\widetilde g_2^1\widetilde g_2^2|/M^2$\\
(GeV$^{-4}$)
\end{tabular}
&
\begin{tabular}{c}
Large-$M$\\
$|\widetilde g_2^1\widetilde g_2^2|/M^2$\\
(GeV$^{-4}$)
\end{tabular}
&
Ref. \\
\hline

$e$--$e$
& $4.08\times10^{-4}$
& ---
& \cite{Terrano2015} \\

$e$--$e^+$
& $5.28\times10^{8}$
& $1.76\times10^{8}$
& \cite{Fadeev2022} \\

$e$--$\mu^+$
& $3.52$
& $1.19$
& \cite{Fadeev2022} \\

$e$--$e$ (He)
& $2.68\times10^{5}$
& ---
& \cite{Fadeev2022} \\

$e$--$\bar p$
& $4.04\times10^{3}$
& ---
& \cite{Ficek2018} \\

$e$--$n$
& $3.73\times10^{-8}$
& ---
& \cite{Almasi2020} \\

$p$--$p$
& $1.87\times10^{-4}$
& ---
& \cite{Ramsey1979} \\

$n$--$n$
& $4.66\times10^{-9}$
& ---
& \cite{Vasilakis2009} \\

\end{tabular}
\end{ruledtabular}
\end{table}

\clearpage 
\subsection{Representative limits}
\label{Appendix2}
The representative limits below are obtained by digitizing published exclusion curves using WebPlotDigitizer and translating selected points through the corresponding curve. For each experiment, a representative point is chosen near the region of strongest sensitivity. These values are approximate and are intended to provide representative constraints rather than a complete translation of the exclusion curves over the full mediator-mass range. Table~\ref{tab:RMPplotguide} provides a guide to the published exclusion curves that can be used, together with the mappings above, to obtain the full mass-dependent constraints.

\begin{table}[b]
\caption{Representative limits on $f_{4+5}$ and the corresponding translated bounds on the spin-1 ($S=1$) vector--tensor coupling, obtained by matching the common $V_{4+5}$ component using Eq.~\eqref{eq:boundVT}. These bounds are component-wise $V_{4+5}$ reinterpretations. The complete vector--tensor interaction also
contains the $\bq^2V_1$ and $V_2+V_3$ contributions, which are not included in this translation and may contribute to the response of a particular experiment.}
\label{tab:S1VT}
\centering\scriptsize
\setlength{\tabcolsep}{2.5pt}
\begin{tabular}{@{}lllll@{}}
\toprule
Pair & Range $\lambda$ (m) & $|f_{4+5}|$ & $|g_T^1g_V^2|$ (GeV$^{-1}$) & Ref.\\
\midrule
$e$--$N$ & $10^{-7}$--$10^{-5}$ & $5.12\times10^{-2}$ & $2.50\times10^{1}$ & \cite{Xiao2024}\\
$e$--$N$ & $2.42\times10^{-5}$ & $4.64\times10^{-8}$ & $2.27\times10^{-5}$ & \cite{Ding2020}\\
$e$--$N$ & $3.92\times10^{-2}$ & $3.40\times10^{-20}$ & $1.66\times10^{-17}$ & \cite{Kim2018}\\
\bottomrule
\end{tabular}
\end{table}

\begin{table}[b]
\caption{Representative laboratory $g_pg_s$ constraints and translated
spin-1 ($S=1$) vector--pseudotensor bounds, checked with Eq.~\eqref{eq:boundVPT}. These bounds are component-wise $V_{9+10}$ reinterpretations.
The complete vector--pseudotensor interaction also contains the momentum-suppressed $\bq^2V_{14,15}$ contributions, which are not included in this translation.}
\label{tab:S1VPT}
\centering\scriptsize
\setlength{\tabcolsep}{2.5pt}
\begin{tabular}{@{}lllll@{}}
\toprule
Pair & Range $\lambda$ (m) & $|g_pg_s|$ & $|g_{\widetilde T}^1g_V^2|$ (GeV$^{-1}$) & Ref.\\
\midrule
$e$--$e$ & $\sim1$ & $1.12\times10^{-31}$ & $5.48\times10^{-29}$ & \cite{Crescini2022}\\
$e$--$N$ & $1.02$ & $3.82\times10^{-32}$ & $1.87\times10^{-29}$ & \cite{Crescini2022}\\
$e$--$N$ & $1.02$ & $8.87\times10^{-29}$ & $4.34\times10^{-26}$ & \cite{Terrano2015}\\
$e$--$N$ & $9.06\times10^{3}$ & $3.90\times10^{-33}$ & $1.91\times10^{-30}$ & \cite{Heckel2008}\\
$n$--$N$ & $1.02\times10^{8}$ & $3.53\times10^{-36}$ & $9.39\times10^{-37}$ & \cite{Zhang2023}\\
$n$--$N$ & $1.02\times10^{8}$ & $6.05\times10^{-35}$ & $1.61\times10^{-35}$ & \cite{Venema1992}\\
\bottomrule
\end{tabular}
\end{table}

\begin{table}[b]
\caption{EDM-derived $g_pg_s$ constraints and the corresponding translated bounds on the spin-1 ($S=1$) vector--pseudotensor coupling, obtained by matching the $V_{9+10}$ component. These bounds are component-wise $V_{9+10}$ reinterpretations. The complete vector--pseudotensor interaction also contains the
momentum-suppressed $\bq^2V_{14,15}$ contributions, which are not
included in this translation.}
\label{tab:S1VPTEDM}
\centering\scriptsize
\setlength{\tabcolsep}{2.5pt}
\begin{tabular}{@{}lllll@{}}
\toprule
Pair & Range $\lambda$ (m) & $|g_p^1g_s^2|$ & $|g_{\widetilde T}^1g_V^2|$ (GeV$^{-1}$) & Ref.\\
\midrule
$e$--$e$ & $1.17\times10^{-8}$ & $3.24\times10^{-18}$ & $1.59\times10^{-15}$ & \cite{Stadnik2018}\\
$e$--$e$ & $5.38\times10^{-9}$ & $1.99\times10^{-20}$ & $9.74\times10^{-18}$ & \cite{Prosnyak2023}\\
$e$--$N$ & $2.20\times10^{-9}$ & $1.48\times10^{-18}$ & $7.24\times10^{-16}$ & \cite{Stadnik2018}\\
$e$--$N$ & $2.25\times10^{-9}$ & $1.13\times10^{-20}$ & $5.53\times10^{-18}$ & \cite{Prosnyak2023}\\
$N$--$e$ & $1.07\times10^{-8}$ & $6.31\times10^{-17}$ & $1.68\times10^{-17}$ & \cite{Dzuba2018}\\
\bottomrule
\end{tabular}
\end{table}

\begin{table}[t]
\caption{Representative $V_{11}$ limits on $g_Ag_V$ and the corresponding translated bounds on the spin-1 ($S=1$)
axial--tensor coupling, obtained by matching the common $V_{11}$ component using Eq.~\eqref{eq:boundAT}.
These bounds are component-wise $V_{11}$ reinterpretations. The complete axial--tensor interaction also contains the
momentum-suppressed $\bq^2V_{12+13}$ contribution, which is not included in this translation.}
\label{tab:S1AT}
\centering\scriptsize
\setlength{\tabcolsep}{2.5pt}
\begin{tabular}{@{}lllll@{}}
\toprule
Pair & Range $\lambda$ (m) & $|g_A^1g_V^2|$ & $|g_A^1g_T^2|$ (GeV$^{-1}$) & Ref.\\
\midrule
$e$--$e$ & $\sim1$ & $2.39\times10^{-28}$ & $1.17\times10^{-25}$ & \cite{Heckel2013}\\
$e$--$e$ & $1.17\times10^{6}$ & $1.13\times10^{-28}$ & $5.53\times10^{-26}$ & \cite{Hunter2013}\\
$e$--$n$ & $9.56\times10^{5}$ & $5.84\times10^{-24}$ & $1.55\times10^{-24}$ & \cite{Hunter2013}\\
$e$--$n$ & $9.45\times10^{-1}$ & $5.68\times10^{-18}$ & $1.51\times10^{-18}$ & \cite{Wang2023}\\
$n$--$p$ & $\sim10^{-1}$ & $5.60\times10^{-18}$ & $1.49\times10^{-18}$ & \cite{Wang2023}\\
\bottomrule
\end{tabular}
\end{table}

\begin{table}[t]
\caption{Representative TOPE $V_{6+7}$ limits and the corresponding translated bounds on the spin-1 ($S=1$) axial--pseudotensor coupling, obtained by matching the common $V_{6+7}$ component.
Eq.~\eqref{eq:f67conventions}
gives the conversion between the legacy experimental $f_{6+7}$ normalization used in the source table and the RMP/DM normalization. The $e$--$n$ entry from Ref.~\cite{Ji2017} is a projected limit from a proposed experiment.}

\label{tab:S1APT}
\centering\scriptsize
\setlength{\tabcolsep}{2.5pt}
\begin{tabular}{@{}lllll@{}}
\toprule
Pair & Range $\lambda$ (m) & $|f_{6+7}|$ & $|g_A^1g_{\widetilde T}^2|$ (GeV$^{-1}$) & Ref.\\
\midrule
$e$--$e$ & $1.02\times10^{1}$ & $7.53\times10^{-18}$ & $1.47\times10^{-14}$ & \cite{Ji2018}\\
$e$--$n$ & $5.84$ & $4.67\times10^{-25}$ & $4.57\times10^{-22}$ & \cite{Ji2017}(projected)\\
\bottomrule
\end{tabular}
\end{table}

\begin{table}[t]
\caption{Representative $g_pg_p$ limits and the corresponding
translated bounds on the spin-1 ($S=1$) tensor--tensor and
pseudotensor--pseudotensor couplings, obtained by matching the
common $V_3$ component using Eq.~\eqref{eq:boundTTpp}.
These bounds are component-wise $V_3$ reinterpretations.
For $g_Tg_T$, the complete tensor--tensor interaction also contains
a $V_2$ contribution, which is not included in this translation.
For $g_{\widetilde T}g_{\widetilde T}$, the interaction is a pure
$V_3$ structure at the order retained.}
\label{tab:S1TT}
\centering\scriptsize
\setlength{\tabcolsep}{2.5pt}
\begin{tabular}{@{}lllll@{}}
\toprule
Pair & Range $\lambda$ (m) & $|g_pg_p|$ & $|g_Tg_T|=|g_{\widetilde T}g_{\widetilde T}|$ (GeV$^{-2}$) & Ref.\\
\midrule
$e$--$e$ & $1.57$ & $5.09\times10^{-17}$ & $1.22\times10^{-11}$ & \cite{Terrano2015}\\
$e$--$e$ & $1.35\times10^{-8}$ & $4.60\times10^{-8}$ & $1.10\times10^{-2}$ & \cite{Fadeev2022}\\
$e$--$e^+$ & $1.61\times10^{-8}$ & $2.66\times10^{-6}$ & $6.37\times10^{-1}$ & \cite{Fadeev2022}\\
$e$--$n$ & $6.07$ & $1.58\times10^{-14}$ & $2.06\times10^{-12}$ & \cite{Almasi2020}\\
$e$--$p$ & $2.26\times10^{-11}$ & $1.17\times10^{-6}$ & $1.52\times10^{-4}$ & \cite{Fadeev2022}\\
$n$--$n$ & $6.6$ & $4.78\times10^{-9}$ & $3.38\times10^{-10}$ & \cite{Vasilakis2009}\\
$p$--$p$ & $1.45\times10^{-6}$ & $2.51\times10^{-4}$ & $1.78\times10^{-5}$ & \cite{Ramsey1979}\\
\bottomrule
\end{tabular}
\end{table}

\begin{table}[b]
\caption{Spin-2 tensor--axial-tensor bounds obtained from the shared
$V_{11}$ component of Table~\ref{tab:S1AT} using
Eq.~\eqref{eq:S2V11bound}.  Spin-2 exchange also generates $V_{12,13}$.}
\label{tab:S2V11}
\centering\footnotesize
\begin{tabular}{llll}
\toprule
Pair & Range $\lambda$ (m) & $|g_{21}\widetilde g_{22}|$ (GeV$^{-2}$) & Ref.\\
\midrule
$g_2^e\widetilde g_2^e$ & $\sim1$ & $1.83\times10^{-21}$ & \cite{Heckel2013}\\
$g_2^e\widetilde g_2^e$ & $1.17\times10^{6}$ & $8.66\times10^{-22}$ & \cite{Hunter2013}\\
$g_2^e\widetilde g_2^n$ & $9.56\times10^{5}$ & $1.32\times10^{-23}$ & \cite{Hunter2013}\\
$g_2^e\widetilde g_2^n$ & $9.45\times10^{-1}$ & $1.29\times10^{-17}$ & \cite{Wang2023}\\
$g_2^n\widetilde g_2^p$ & $\sim10^{-1}$ & $1.27\times10^{-17}$ & \cite{Wang2023}\\
\bottomrule
\end{tabular}
\end{table}
\clearpage
\begin{table}[h]
\caption{Spin-2 axial-tensor--axial-tensor bounds from the common $V_3$
component of the spin-1 pseudotensor--pseudotensor interaction in
Table~\ref{tab:S1TT}.  Equation~\eqref{eq:S2PTPTbound} is used; the additional
spin-2 $V_2$ term is not included in this translation.}
\label{tab:S2PTPT}
\centering\footnotesize
\begin{tabular}{llll}
\toprule
Pair & Range $\lambda$ (m) & $|\widetilde g_{21}\widetilde g_{22}|$ (GeV$^{-2}$) & Ref.\\
\midrule
$e$--$e$ & $1.57$ & $5.90\times10^{-36}$ & \cite{Terrano2015}\\
$e$--$e$ & $1.35\times10^{-8}$ & $7.21\times10^{-11}$ & \cite{Fadeev2022}\\
$e$--$e^+$ & $1.61\times10^{-8}$ & $2.93\times10^{-9}$ & \cite{Fadeev2022}\\
$e$--$n$ & $6.07$ & $3.62\times10^{-41}$ & \cite{Almasi2020}\\
$e$--$p$ & $2.26\times10^{-11}$ & $1.94\times10^{-10}$ & \cite{Fadeev2022}\\
$n$--$n$ & $6.6$ & $2.74\times10^{-42}$ &  \cite{Vasilakis2009}\\
$p$--$p$ & $1.45\times10^{-6}$ & $3.00\times10^{-24}$ & \cite{Ramsey1979}\\
\bottomrule
\end{tabular}
\end{table}

\begin{table}[h]
\caption{Spin-2 $g_2g_2$ bounds obtained from the $V_3$-based
$g_Tg_T$ limits in Table~\ref{tab:S1TT}, using
$|g_2^1g_2^2|=32|g_T^1g_T^2|$.
This is a component-wise translation obtained by matching the
$V_3$ term. The additional $V_2$ term accompanying $V_3$, as well
as the $V_1$ and $V_{4,5}$ terms are not included in this translation.}
\label{tab:S2TT}
\centering\footnotesize
\begin{tabular}{llll}
\toprule
Pair & Range $\lambda$ (m) & $|g_{21}g_{22}|$ (GeV$^{-2}$) & Ref.\\
\midrule
$e$--$e$ & $1.57$ & $3.90\times10^{-10}$ & \cite{Terrano2015}\\
$e$--$e$ & $1.35\times10^{-8}$ & $3.52\times10^{-1}$ & \cite{Fadeev2022}\\
$e$--$e^+$ & $1.61\times10^{-8}$ & $2.04\times10^{1}$ & \cite{Fadeev2022}\\
$e$--$n$ & $6.07$ & $6.58\times10^{-11}$ & \cite{Almasi2020}\\
$e$--$p$ & $2.26\times10^{-11}$ & $4.88\times10^{-3}$ & \cite{Fadeev2022}\\
$n$--$n$ & $6.6$ & $1.08\times10^{-8}$ & \cite{Vasilakis2009}\\
$p$--$p$ & $1.45\times10^{-6}$ & $5.70\times10^{-4}$ & \cite{Ramsey1979}\\
\bottomrule
\end{tabular}
\end{table}
 
\section{Nonrelativistic fermion bilinears}
\label{app:NR}

With $q=p'-p$, $\bp=(\bp^{\rm in}+\bp^{\rm out})/2$, and $\bar uu\simeq2m$, the leading bilinears are
\begin{align}
\bar u'\gamma^0u&\simeq2m,
\label{eq:NRV0}\\
\bar u'\boldsymbol\gamma u&\simeq2\bp-i\bq\times\bsig,
\label{eq:NRVi}\\
\bar u'\gamma^0\gamma_5u&\simeq2\bsig\cdot\bp,
\label{eq:NRA0}\\
\bar u'\boldsymbol\gamma\gamma_5u&\simeq2m\bsig,
\label{eq:NRAi}\\
\bar u'\gamma_5u&\simeq-\bsig\cdot\bq.
\label{eq:NRP}
\end{align}
For the contracted tensor current,
\begin{align}
T^0&\equiv q_\alpha\bar u'\sigma^{\alpha0}u
\simeq-i\bq^2+2\bsig\cdot(\bp\times\bq),
\label{eq:NRT0}\\
T^i&\equiv q_\alpha\bar u'\sigma^{\alpha i}u
\simeq2m(\bq\times\bsig)^i,
\label{eq:NRTi}
\end{align}
and for the contracted pseudotensor current,
\begin{align}
\widetilde T^0&\equiv q_\alpha\bar u'\sigma^{\alpha0}\gamma_5u
\simeq2im(\bsig\cdot\bq),
\label{eq:NRPT0}\\
\widetilde T^i&\equiv q_\alpha\bar u'\sigma^{\alpha i}\gamma_5u
\simeq2i\,p^i(\bsig\cdot\bq).
\label{eq:NRPTi}
\end{align}
At the second vertex the momentum entering the fermion line is $-q$, producing the corresponding sign reversals.  The exact Gordon identity
\begin{equation}
q_\mu\bar u'\sigma^{\mu\nu}u=
 i\bar u'(2m\gamma^\nu-K^\nu)u
\end{equation}
provides a useful check on Eqs.~\eqref{eq:NRT0} and \eqref{eq:NRTi}.

\section{Distributional Fourier transforms}
\label{app:FT}

The basic identities used throughout the paper are
\begin{align}
\int\frac{\dd^3q}{(2\pi)^3}
\frac{e^{i\bq\cdot\br}}{\bq^2+M^2}&=Y_M(r),
\label{eq:FT1}\\
\int\frac{\dd^3q}{(2\pi)^3}e^{i\bq\cdot\br}
\frac{\bq^2}{\bq^2+M^2}&=D_M(\br),
\label{eq:FT2}\\
\int\frac{\dd^3q}{(2\pi)^3}e^{i\bq\cdot\br}
\frac{q_iq_j}{\bq^2+M^2}&=H_{ij}(\br).
\label{eq:FT3}
\end{align}
The contact term in Eq.~\eqref{eq:H} follows from the distributional identity
\begin{equation}
\partial_i\partial_j\frac1r=
\frac{3\hat r_i\hat r_j-\delta_{ij}}{r^3}
-\frac{4\pi}{3}\delta_{ij}\delta^3(\br).
\end{equation}
These terms must be retained when the interaction range is comparable to or shorter than atomic length scales.

\section{Comment on Hermiticity and complex couplings}
\label{app:hermiticity}

For diagonal interactions of a neutral Hermitian boson, it is convenient to choose a basis of Hermitian fermion operators and real coefficients.  In particular,
\begin{equation}
\bar\psi\sigma^{\mu\nu}\psi
\ \text{is Hermitian},\qquad
\bar\psi\sigma^{\mu\nu}\gamma_5\psi
\ \text{is anti-Hermitian},
\end{equation}
so the pseudotensor term is written as
$i g_{\widetilde T}\bar\psi\sigma^{\mu\nu}\gamma_5\psi$ with real $g_{\widetilde T}$.  Equivalently one may omit the explicit $i$ and use a purely imaginary coefficient.  These are the same Hermitian interaction written in two conventions.  A physical $T$-violating phase arises only when a phase cannot be removed by such an operator redefinition, for example in an appropriate non-diagonal coupling matrix.  The $T$-odd spin-1 potentials in Sec.~\ref{sec:spin1} arise from the transformation properties of the Hermitian operators themselves and do not require complex diagonal couplings.

\end{document}